\documentclass[aps,showpacs,floatfix,superscriptaddress]{revtex4}

\usepackage{amsmath,fixmath}
\usepackage{acronym}
\usepackage{amssymb}
\usepackage{amsfonts}
\usepackage{color}
\usepackage{bm}
\usepackage{feynmp}
\usepackage[pdftex]{graphicx}
\usepackage{ifpdf}
\ifpdf
\else
\fi

\newcommand{\risz}[1]{G_0 (\vn p,z) #1}

\def\vm#1{\boldsymbol{#1}}
\def\vn#1{\textbf{#1}}
\def\mat{\mathcal{M}}

\acrodef{MCT}{mode-coupling theory}
\acrodef{ERM}{Euclidean random matrices}
\acrodef{DOS}{density of states}

\begin{document}

\title{On the high-density expansion for
  Euclidean Random Matrices}

\author{T.~S.~Grigera} \affiliation{ Instituto de Investigaciones
  Fisicoqu\'{\i}micas Te\'oricas y Aplicadas (INIFTA) and Departamento
  de F\'{\i}sica, Facultad de Ciencias Exactas, Universidad Nacional
  de La Plata, c.c. 16, suc. 4, 1900 La Plata, Argentina}

\affiliation{CCT La Plata, Consejo Nacional de Investigaciones
  Cient\'{\i}ficas y T\'ecnicas, Argentina}

\author{V.~Martin-Mayor} \affiliation{Departamento de F\'\i{}sica
  Te\'orica I, Universidad Complutense, 28040 Madrid, Spain.}

\affiliation{Instituto de Biocomputaci\'on y
  F\'{\i}sica de Sistemas Complejos (BIFI), Zaragoza, Spain.}

\author{G.~Parisi} \affiliation{Dipartimento di Fisica, INFM and
  INFN, Universit\`a di Roma ``La Sapienza'', 00185 Roma, Italy.}

\author{P.~Urbani} \affiliation{Dipartimento di Fisica, Universit\`a
  di Roma ``La Sapienza'', 00185 Roma, Italy.}

\author{P. Verrocchio}
  \affiliation{Dipartimento di Fisica, Universit{\`a} di Trento, via Sommarive 14, 38050 Povo, Trento,
  Italy.}
\affiliation{Istituto Sistemi Complessi (ISC-CNR),
    Via dei Taurini 19, 00185 Roma, Italy}

\begin{abstract}
  Diagrammatic techniques to compute perturbatively the spectral
  properties of \acf{ERM} in the high-density regime are introduced
  and discussed in detail. Such techniques are developed in two
  alternative and very different formulations of the mathematical
  problem and are shown to give identical results up to second order
  in the perturbative expansion. One method, based on writing the
  so-called resolvent function as a Taylor series, allows to group the
  diagrams in a small number of topological classes, providing a
  simple way to determine the infrared (small momenta) behavior of the
  theory up to third order, which is of interest for the comparison
  with experiments. The other method, which reformulates the problem
  as a field theory, can instead be used to study the infrared
  behaviour at any perturbative order.
\end{abstract}


\maketitle

\section{Introduction}

Random matrices~\cite{MEHTA} are $N\times N$ matrices whose entries
are random numbers drawn from a certain probability distribution.
Their statistical spectral properties in the large $N$ limit describe
a wide range of physical phenomena: nuclear spectra~\cite{WIGNER},
quantum chaos~\cite{BOHIGAS}, localization in electronic
systems~\cite{LOCALIZATION}, diffusion in random
graphs~\cite{RANDOM-GRAPHS}, liquid dynamics~\cite{INM} and the glass
transition~\cite{GLASSES}, complex networks~\cite{NETWORKS},
superstrings~\cite{SUPERSTRINGS}. Random matrices may be grouped in a
few universality classes according to their statistical
properties~\cite{MEHTA}. For most of these classes, the density of
eigenvalues follows Wigner's semicircle law. It has thus become of
interest to identify ensembles where the semicircle law is modified in
a non-trivial way. One such ensemble results when the
corresponding physical problem has a conserved quantity (e.g. momentum
in case of propagating excitations, or number density in diffusion
problems).  Under such circumstances, the random matrix that best
describes the problem is typically a Laplacian
matrix~\cite{RANDOM-GRAPHS}, which has the property
\begin{equation}
\sum_j {\cal M}_{ij} = 0.
\label{ZERO}
\end{equation}
This encodes the property that a vector whose components are identical
is an eigenvector with eigenvalue zero.

A kind of random matrices of particular relevance in the study of
off-lattice systems are the so-called
\acf{ERM}~\cite{MPZ,MATHREF}. Place $N$ particles in positions
$\vn{x}_i$, $i=1,2,\ldots, N$, belonging to some region of
$D$-dimensional Euclidean space, of volume $V$. The positions are
drawn randomly from some probability distribution function
$P(\{\vn{x}_i\})$. The entries of an ERM are a deterministic function
of these random positions, $M_{ij}=f(\vn{x}_i-\vn{x}_j)$. If a
conservation law is relevant for the problem at hand, we will rather
have a Laplacian \ac{ERM}:
\begin{equation}
{\cal M}_{ij}=\delta_{ij} \sum_{k=1}^N f(\vn{x}_j-\vn{x}_k)\ -\
f(\vn{x}_i-\vn{x}_j). 
\label{DEFERM}
\end{equation}
Note that we never find the same particle label twice in the argument
of the function $f(\vn{x}_i-\vn{x}_j)$, since the term
$f(\vn{x}_i-\vn{x}_i)$ cancels. In a diagonal term,
$\delta_{ij}f(\vn{x}_i-\vn{x}_k)$, the $k$th particle shall be called
a {\em medium} particle, while the $i$th particle will be the \emph{chain}
particle.

The function $f$ in Eq.~(\ref{DEFERM}) is quite general: only
rotational invariance and the existence of the Fourier transform $\hat
f(p)$  are
assumed ($p=\sqrt{\vn{p}\cdot\vn{p}}$). Furthermore, even if in this work $f$ will be a scalar
function, for some applications it should rather be a matrix-valued
function. It must be so, for instance, to account for the vector
nature (longitudinal or transversal) of vibrational
dynamics~\cite{ERM4}. Most of our results extend as such to this more
general case.

\acp{ERM} describe {\em topologically disordered systems}, at variance
with problems were the $N$ positions $\{\vn{x}_i\}$ are placed on a
crystalline lattice~\cite{DIS-CRYSTAL}. We will be considering a
extreme case, in which the N positions are placed with uniform
probability on the volume $V$.  The particle-number density,
$\rho=N/V$ will be held fixed while we take the large $N$ limit. Note
that there are two sources of statistical correlation among the
entries of matrix (\ref{DEFERM}), even if the positions $\{\vn{x}_i\}$
are totally uncorrelated. First, it is a Laplacian matrix, recall
Eq.~(\ref{ZERO}). Second, due to the triangular inequality of
Euclidean geometry, the distances from two neighbouring particles to a
third one are necessarily similar.

Specific applications of \acp{ERM} include disordered $d$-wave
superconductors~\cite{CHAMON}, disordered magnetic
semiconductors~\cite{BRGS03} (very similar to a spin-glass
model~\cite{dean}), Instantaneous Normal Modes in
liquids~\cite{INM,cagi00}, vibrations in
glasses~\cite{GLASSES,ERM1,ERM2,ERM3,ERM4}, the gelation transition in
polymers~\cite{BRODERIX} and vibrations in DNA~\cite{COCCO}. \acp{ERM}
have been studied analytically and numerically both in the low
particle-number density regime~\cite{cagi00,BiMo,dean02} and for high
densities~\cite{MPZ,ERM1,ERM2,ERM3,ERM4,SCHIRMACHER}.

In order to compute the basic spectral properties of \ac{ERM} it turns
out to be convenient to introduce the \emph{resolvent,}
\begin{equation}
G(\vn{p}, z)= \lim_{N\to\infty} \overline{\frac{1}{N}\sum_{i,j=1}^N \mathrm{e}^{{\mathrm
      i} \vn{p}\cdot(\vn{x}_j -\vn{x}_i)}
  \left[\frac{1}{z-\mat}\right]_{ij}}\,,
\label{RESOLVENT}
\end{equation}
where the complex number $z \equiv \lambda + \mathrm{i} \eta$ has a
tiny imaginary part $\eta$ and the overline stands for an
average over the $\{\vn{x}_i\}$. If the \ac{ERM} describes physical
excitations (phonons, electrons, etc.) in topologically disordered
systems, the resolvent (\ref{RESOLVENT}) corresponds to the
single-particle Green function, or propagator, for such excitations.
If the system is isotropic, the resolvent depends only on $p$.

The density of eigenvalues $g(\lambda)$, or \acf{DOS}, is given by
\begin{equation}
g(\lambda)= -\frac{1}{\pi} \lim_{p\to\infty} \mathrm{Im} [G(\vn p, \lambda
+\mathrm{i} 0^+)]. \label{DOS-DEF}
\end{equation}
This limiting behaviour is characteristic of topologically disordered
systems. It does not hold for lattice systems.  We note as well that
the constraint (\ref{ZERO}) implies that a plane wave
$\mathrm{e}^{\mathrm{i} \vn{p}\cdot\vn{x}_i}$ is an eigenvector
of the matrix (\ref{DEFERM}) if $p=0$:
\begin{equation}
G(\vn 0, z)=\frac{1}{z}.\label{EXACT-p-to-zero}
\end{equation}

As we shall discuss below, the resolvent takes a very simple form in
the high-density limit (it is actually the \emph{bare} propagator of
the theory):
\begin{equation}
G_0(\vn p,z)=\frac{1}{z-\epsilon(\vn p)},\quad \epsilon(\vn p)=\rho[\hat f(\vn 0)-\hat f(\vn p)].
\label{DEF:G0}
\end{equation}
The physical interpretation is quite appealing~\cite{MPZ}. The system
behaves as an elastic continuum medium. In the large $\rho$ limit, the
plane waves $\mathrm{e}^{\mathrm{i} \vn{p}\cdot\vn{x}_i}$ become exact
eigenvectors of the matrix (\ref{DEFERM}), with eigenvalues given by
the dispersion relation $\epsilon(\vn p)=\rho[\hat f(\vn 0)-\hat f(\vn p)]$. In
particular, for small $p$, $\epsilon(\vn p)= c^2 p^2+{\cal O}(p^4)$, where
$c$ is the \emph{speed of sound}. This neat physical picture motivates
the introduction of a high-density expansion.

At large, but finite $\rho$, the resolvent can be written 
\begin{equation}
G(\vn p,z)=\frac{1}{z-\epsilon(\vn p)-\Sigma(\vn p,z)}\,.
\label{G:DEF}
\end{equation}
The \emph{self-energy} $\Sigma(\vn p,z)$, which is introduced to encode
all the information about the {\em interactions} (a standard practice
in the Green function formalism), vanishes when $\rho$ tends to
$\infty$. In our case, the interaction involved is that between the
propagating excitations and the topological disorder.  An important
theoretical challenge is to compute the self-energy at finite
densities $\rho$. In fact, in this case an eigenvector can be thought
as a packet of plane waves (see section~\ref{SECT:EXAMPLE}). The width
of such packet is related to the imaginary part of $\Sigma$.

Some of us have argued that in the limit of small $p,z$ the leading
term at $1/\rho^2$ order has the form~\cite{ERM1,ERM2,ERM3}
\begin{equation}
\text{Im} \Sigma(\vn p,z+i0^+) ={\cal A}\,
z^{\frac{D-2}{2}} p^2\ +\ {\cal O}(z^{\frac{D-2}{2}}p^4\,,\, z^D p^2)\,,
\label{WRONG}
\end{equation}
($D$ is the space dimension, while ${\cal A}$ is an amplitude). This
has been disputed recently by an independent computation to order
$1/\rho^2$, claiming that the actual analytic structure of the
self-energy for small $z$ and $p$ is~\cite{SCHIRMACHER}
\begin{equation}
\text{Im} \Sigma(\vn p, z+i0^+) \propto
z^{\frac{D}{2}} p^2 + \ldots,
\label{WRONG2}
\end{equation}
analogous to what one finds in the Rayleigh theory of scattering and
in lattice models where disordered spring constants mimick the effect
of topological disorder~\cite{DIS-CRYSTAL}.

By reconsidering in detail the perturbative expansion, in this work we
show that the prefactor ${\cal A}$ in Eq.~(\ref{WRONG}) is actually
null, due algebraic cancellations, and that this cancellations arise
at all orders in the perturbative expansion in $1/\rho$. This is not
related to any known symmetry of the problem, but rather reflects the
mathematical structure of the perturbative contributions. On the other
hand, we will also show that the result in
Ref.~\onlinecite{SCHIRMACHER}, recall Eq.~(\ref{WRONG2}), is
incomplete, since the imaginary part admits a formal expansion for
small $z$
\begin{equation}
\text{Im} \Sigma(\vn p,z+i0^+) =
z^{\frac{D-2}{2}}\sum_{n=0}^\infty g_n(p^2)\,
z^{n}\,,\label{ASYMPTOTICS}
\end{equation}
The constraint (\ref{EXACT-p-to-zero}) implies that $g_n(0)=0$ for all
$n$, so that in general $g_n(p)= {\cal A}_n p^2 + {\cal
  O}(p^4)$. However, we find that, for all functions $f$ and all
$\rho$, ${\cal A}_0=0$, so that $g_0(p) \sim p^4$ while $g_1(p) \sim
p^2$. In this respect, we confirm that the interaction between free
excitations and disorder in topologically disordered systems (as long
as \acp{ERM} describe them) has a peculiar mathematical structure that
is different from disordered lattice systems (for lattice systems
$g_0$ vanishes identically).

To show this we shall compute the self-energy perturbatively within
two unrelated approaches: a) an improved form of the combinatorial
formalism introduced in~\cite{ERM2}, and b) a field-theoretic
formulation. The field theory introduced here is quite different from
standard formalisms in the theory of Random Matrices (see e.g.
Refs.~\cite{MPZ,ERM1}). It probably deserves an indepth study, which
is left for future work. We remark that our combinatorial formalism is
simpler than the field theory, and is probably the method of choice to
carry out higher-order computations in the $1/\rho$
expansion. However, it has the drawback that the asymptotic
$g_0(p^2)\sim p^4$ appears at order $1/\rho^2$ from an exact
cancellation of two contributions of order $p^2$ (at order $1/\rho^3$
we find an exact cancellation of \emph{ten} contributions of order
$p^2$). The field-theoretic framework clarifies that these
cancellations are not accidental, and thus not restricted to low
orders in the $1/\rho$ expansion.

The layout of the remaining part of this work is as follows: in
sec.~\ref{SECT:EXAMPLE} we discuss a particular
phenomenon (phonons in topologically disordered systems) where a
theory based on \acp{ERM} has been proposed in recent years. In
sec.~\ref{SECT:MAIN} we anticipate our main result, namely the leading
order of $\text{Im} \Sigma(\vn p,z+i0^+)$. In sec.~\ref{SECT:COMBINATORIAL} we
discuss in detail the combinatorial formalism up order 
$1/\rho^2$. We describe the rules to group all the diagrams that
arise at this order in a very small number of diagrams, according to
their topological structure, and show that up to second order in the
function $g_0(p^2)$ the prefactor of the term $\propto p^2$ cancels
out. We also see that this cancellation appears in a given class of
diagrams at $1/\rho^3$.  In order to shed a light over the
mathematical origin of such cancellation, in
sec.~\ref{FIELD} we introduce a field-theoretical
formulation that, despite producing a much larger number of diagrams,
allows to give an argument explaining the origin of the cancellation at
any perturbative order.

\section{A case study for \ac{ERM}: phonons in topologically
  disordered systems}\label{SECT:EXAMPLE}

Although \acp{ERM} have a wide range of application, in this paper we
are mainly interested in the study of phonons in amorphous systems, such
as glasses or supercooled liquids~\cite{HFS}, since the big amount of
experimental evidences may provide fundamental insights about the
correctness of the theory. Of particular interest is the case where
the frequencies $\omega(p)$ of the phonons with wave vector $p$
lie in the GHz to the THz region (high-frequency sound).  This
is in fact the range explored by neutron and $X$-ray inelastic
scattering experiments. These give the inelastic contribution to the
dynamic structure factor, i.e.\ a Brillouin-like peak with position
$\omega(p)$ and width $\Gamma(p)$. Summarizing the experimental
findings, for $p < p_0$ ($p_0$ is the first maximum of the static
structure factor, typically a few nm$^{-1}$~\cite{HANSEN}) one finds a
linear dispersion relation $\omega(p) \sim cp$, where the speed of
sound $c$ is quite close to that obtained by acoustic
measurements. The dispersion relation typically saturates at $p\sim
p_0$.  Moreover, the $p$-dependence of the peak width is often
described by $\Gamma(p) \propto p^\alpha$. Interestingly enough,
$\Gamma(p)$ also saturates as the momentum becomes $p\sim p_0$. There
has been a hot debate among different experimental groups
about the value of the exponent $\alpha$~\cite{POLEMICA}, some
claiming $\alpha \sim 2$, and some $\alpha \sim 4$. There is now some
consensus that in the region where $\Gamma$ is
independent of temperature (i.e.\ $\omega(p) \geq 1\,$THz) one has
$\alpha=4$, while at lower frequencies (the GHz region), where
$\Gamma$ has a strong temperature dependence, the experimental value
is $\alpha=2$~\cite{POLEMICARISOLTA}.

A simple model of the high-frequency sound is afforded by scalar
harmonic vibrations around a topologically disordered structure made
of $N$ oscillation centers $\vn{x}_i$, placed with uniform probability
on a volume $V$\footnote{One may just as easily introduce a model for
  vector harmonic vibrations~\cite{ERM4}. In fact, all our results
  within the combinatorial formalism translate directly to the vector
  case.}. Particle displacements $\varphi_i$ have an elastic energy
\begin{equation}
U\bigl(\{\varphi_i\}\bigr) = \frac{1}{2}\sum_{i,j=1}^N f(\vn{x}_i-\vn{x}_j)
(\varphi_i-\varphi_j)^2= 
\sum_{i,j=1}^N {\cal M}_{i,j} \varphi_i\varphi_j,\label{EQ:M-DEF} 
\end{equation}
where the matrix ${\cal M}$ has the form Eq.~(\ref{DEFERM})
and $f(\vn{x})$ is the spring constant connecting particles
separated by the vector $\vn{x}$. We assume that $f(\vn{x})$ is
spherically symmetric, so that $\hat f(\vn{p})=g(p^2)$, where $g$ is a
smooth function. In the framework of the one-phonon approximation, 
the inelastic dynamic structure factor is related to the
resolvent via
\begin{equation}
S(\vn p,\omega) = -\frac{p^2}{\omega \pi}\mathrm{Im} G(\vn p,\omega^2+\mathrm{i} 0^+).
\label{DSF}
\end{equation}
As a consequence, the width of the Brillouin peak is related to the
imaginary part of $\Sigma$ by
\begin{equation}
\text{Im} \Sigma(\vn p,\omega(p)) = \omega(p) \Gamma(p).
\label{GAMMA}
\end{equation}
Then Eq.~(\ref{ASYMPTOTICS}) implies that $\Gamma(p) \sim p^4$ for very
small $p$ (for $p\sim p_0$ the width saturates and a mixed, more
complex scaling should be expected). In that regime the
phonon-disorder interaction can be thought of as a scattering
phenomenon of the Rayleigh type.

Since \acp{ERM} describe the dynamics of vibrating particles within
the context of the harmonic approximation, the theoretical predictions
based on \ac{ERM} theory must be compared with experiments in the
region where $\Gamma$ is independent of temperature; in fact the
temperature dependence is an indication that the width of the peak is
rather due to thermal processes, such as anharmonicities or
relaxations, which require more refined theoretical approaches.

We finally mention that vibrational frequencies $\omega$ are related
to \acp{ERM} eigenvalues $\lambda$ ($z=\lambda+\mathrm{i} 0^+$), by
the relation $\lambda=\omega^2$, see Eq.~(\ref{DSF}). Hence, the width
of spectral peaks in $\lambda$-space and in $\omega$-space are related
by Eq.~(\ref{GAMMA}). Furthermore, Eqs.~(\ref{DOS-DEF})
and~(\ref{ASYMPTOTICS}) imply that the \ac{DOS} in $\lambda$ space
behaves for small $\lambda$ as
$g_\lambda(\lambda)\propto\lambda^{(D-2)/2}$, which translates to
frequency space as a Debye spectrum $g_\omega(\omega)\propto
\omega^{D-1}$ (because of the Jacobian in the change of variable:
$\mathrm{d}\lambda=2\omega\mathrm{d}\omega$). At this point, the
reader may object that lattice systems have a Debye spectrum even if
$g_0$ in Eq.~(\ref{ASYMPTOTICS}) vanishes for them. In fact, their Debye
spectrum is possible because Eq.~(\ref{DOS-DEF}) does not hold in
the lattice case.

\section{The main result}\label{SECT:MAIN}

The main result of this work is the following. Expanding the
self-energy in powers of $1/\rho$, i.e.
\begin{equation}
\Sigma(\vn{p},z)=\Sigma^{(1)}(\vn{p},z)+\Sigma^{(2)}(\vn{p},z)+\ldots,
\label{EC:EXPANSION}
\end{equation}
where $\Sigma^{(k)}$ is of order $1/\rho^k$, one has only one first-order
contribution,
\begin{equation}
\Sigma^{(1)}(\vn{p},z)=\frac{1}{\rho}\int\frac{\mathrm{d}^D
  \vn{q}}{(2\pi)^D}\, V(\vn{q},\vn{p}) G_0(\vn{q},z)
V(\vn{q},\vn{p}),
\label{SIGMA1}
\end{equation}
while at second order
\begin{equation}
\Sigma^{(2)}(\vn{p},z)=\Sigma_A^{(2)}(\vn{p},z)+\Sigma_B^{(2)}(\vn{p},z)+\Sigma_B^{(2)}(\vn{p},z),\label{EQ:SIGMA2-TRES-PARTES}
\end{equation}
where the three topologically different pieces are
\begin{subequations}
\label{EQ:SIGMA2-GRUPO}
\begin{align}
\Sigma_A^{(2)}(\vn{p},z) &=\frac{1}{\rho}\int\frac{\mathrm d^D
  \vn{q}}{(2\pi)^D}\, V(\vn{q},\vn{p})
G_0(\vn{q},z)\Sigma^{(1)}(\vn{q},z) G_0(\vn{q},z)
V(\vn{q},\vn{p}),\label{EQ:SIGMAA}\\
\Sigma_B^{(2)}(\vn{p},z) &= \frac{1}{\rho^2}\int
\frac{\mathrm{d}^D \vn{q}}{(2\pi)^D}\frac{\mathrm{d}^D
  \vn{k}}{(2\pi)^D}\, V(\vn{p}-\vn{q},\vn{p}) G_0(\vn{q},z) V(\vn{q}
-\vn{k},\vn{q}) G_0(\vn{k},z)
V(\vn{p}-\vn{q},\vn{p}-\vn{q}+\vn{k})\times\nonumber\\
  &\quad G_0(\vn
  {p}-\vn{q}+\vn{k},z) V(\vn{q}-\vn{k},\vn{p}),\label{EC:SIGMAB}\\
 \Sigma_C^{(2)}(\vn{p},z) &=
\frac{1}{\rho^2}\int \frac{\mathrm{d}^D
  \vn{q}}{(2\pi)^D}\frac{\mathrm{d}^D \vn{k}}{(2\pi)^D}\,
V(\vn{p}-\vn{q},\vn{p}) G_0(\vn{q},z) V(\vn{q}-\vn{k},2\vn{q}-\vn{p})
G_0 (\vn{k},z) V(\vn{p}-\vn{k},\vn{p}).\label{EQ:SIGMA_C}
\end{align}
\end{subequations}
In Eqs.~(\ref{SIGMA1})--(\ref{EQ:SIGMA2-GRUPO}) we have used
\begin{equation}
V(\vn{q},\vn{p})=\rho[\hat f(\vn{q})-\hat f(\vn{p} - \vn{q})],
\label{VERTEX}
\end{equation}
which, as we will see below, plays the role of the interaction
vertex. The bare propagator $G_0$ was defined in
Eq.~(\ref{DEF:G0}). Note that $V(\vn{q},\vn{p})\neq
V(\vn{p},\vn{q})$. Other useful identities are
\begin{equation}
V(\vn{q},\vn{p})=V(-\vn{q},-\vn{p}),\qquad V(\vn{q},\vn{p})=-V(\vn{p}
-\vn{q},\vn{p}).
\label{IDENTITIES}
\end{equation}
Note that since $V(\vn{q},\vn 0)=0$, we have
\begin{equation}
\Sigma_A^{(2)}(\vn 0,z)=\Sigma_B^{(2)}(\vn 0,z)=\Sigma_C^{(2)}(\vn 0,z)=0.
\end{equation}

The high-density expansion for Laplacian \ac{ERM} was introduced
in~\cite{ERM1,ERM2}.  Eq.~(\ref{SIGMA1}) was already reported
there but, instead of Eq.~(\ref{EQ:SIGMA2-TRES-PARTES}), one had $39$
diagrams of order $1/\rho^2$. Even if the final expressions were
cumbersome, a numerical evaluation of the amplitude ${\cal A}$ in
Eq.~(\ref{WRONG}) was attempted for a simple choice of the function
$f$. Presumably because of a numerical mistake, it was wrongly
concluded that ${\cal A}\neq 0$. Afterwards, it was announced (without
supporting technical details) that the 39 diagrams previously found at
order $1/\rho^2$ could be grouped as in
Eq.~(\ref{EQ:SIGMA2-TRES-PARTES})~\cite{ERM4}. Unfortunately, a
numerical reevaluation of the amplitude ${\cal A}$ was not attempted
from these simpler expressions.

We remark as well that an independent computation of $\Sigma$ to order
$1/\rho^2$ has appeared recently~\cite{SCHIRMACHER}. We have checked
that their results are consistent with ours, letting aside contact
terms (in fact, these authors explictly state that some contact terms
are lacking from their final expressions). Thus, their failure in
identifying the $g_0$ term in Eq.~(\ref{ASYMPTOTICS}) is not due to
discrepancies in the final expressions. The underlying reason is
rather more mundane, as we explain below.

At first order the theory has the following behaviour. For small
$\lambda$, $z=\lambda+\mathrm{i} 0^+$, we approximate the imaginary
part of the propagator $G_0$ by
\begin{equation}
\text{Im} G_0(\vn{q},\lambda+\mathrm{i} 0^+)=-\frac{\pi}{2\sqrt
  \lambda}\delta \left(q-\frac{\sqrt \lambda}{c}\right)
\end{equation} 
(assuming a linear dispersion relation $\epsilon(\vn p) \approx c^2 p^2$).
Then the only contribution to the imaginary part comes from
$q=\sqrt{\lambda}/c$. To evaluate the vertex $V(\vn{q},\vn{p})$ at
small $q$, and small $p$ we just need to recall that $\hat f(\vn{p})=
g(p^2)$. It is important to avoid any assumptions about the ratio
$p/q$, which can be either very large or very small when both $p$ and
$q$ are small (at the Brillouin peak $p/q\sim 1$, but in
Ref.~\cite{SCHIRMACHER} it was unjustifiedly assumed that $p \ll
q$). Then
\begin{align}
V(\vn{q},\vn{p})&= g(q^2) - g(q^2+p^2 - 2\vn{p}\cdot\vn{q}) \nonumber\\
&\approx g(0)+g'(0) q^2 - g(0) - g'(0) [ q^2+p^2 -
2\vn{p}\cdot\vn{q}]\nonumber\\
&=- g'(0) [ p^2 - 2\vn{p}\cdot\vn{q}].
\end{align}
If we now square the vertex function and perform the angular integral,
we obtain ($S_D$ is the surface of the sphere in $D$ dimensions)
\begin{equation}
[g'(0)]^2 S_D\left(p^4 +\frac{1}{D} q^2 p^2\right).
\end{equation}
The integral over $q$ is now straightforward, thanks to Dirac's
$\delta$ function in Eq.~(\ref{SIGMA1}). We get
\begin{equation}
\text{Im} \Sigma^{(1)}(\vn{p}, \lambda +{\mathrm i} 0^+)\propto 
-\left[\lambda^{\frac{D-2}{2}}p^4 + \lambda^{D/2} \frac{p^2}{D c^2} \right].
\end{equation}
Hence, already at order $1/\rho$, $g_0(p^2)$ in
Eq.~(\ref{ASYMPTOTICS}) is of order $p^4$. Had we neglected the $p^4$
term in confront of the $q^2 p^2$ term (as done in
Ref.~\cite{SCHIRMACHER}), we would have failed in identyfing the $g_0$
term. The physical reason for which the presence of such a term is
mandatory (namely the existence of a Debye spectrum), was discussed in
the concluding paragraph of Sect.~\ref{SECT:EXAMPLE}.

Let us now check that the amplitude ${\cal A}$ in Eq.~(\ref{WRONG})
vanishes.  We merely need to compute the imaginary part of the
self-energy at its lowest order in $\lambda$, namely
$\lambda^{\frac{D-2}{2}}$. A general term of the diagrammatic
expansion involves the factor
\begin{equation}
\int \!\! \mathrm d^D\!\vn{q}\, G_0(\vn q,z)= \int\!\!\mathrm dq \mathrm d\Omega_D\, q^{D-1}G_0(\vn q,z),
\end{equation}
where we have expressed the measure in terms of polar coordinates in $D$
dimensions. Every bare propagator is associated to one or more
vertices that are smooth functions of the involved momenta. In fact we
can expand the product of such vertices in a Taylor series. Now, the
point is that if we want the lowest order in $z$, we have to exclude
all the terms that are proportional to $q$ and we have to take only
the zeroth order term of the Taylor expansion. We can obtain this term
simply making the following substitution
\begin{equation}
\tilde {\text{Im}} \int \mathrm d^D\vn q\,G_0(\vn q,\lambda+\mathrm{i}
0^+)=-\frac{\pi}{2}\lambda^{(D-2)/2}\int \mathrm dq\,\mathrm d\Omega_D\delta(q)
\end{equation} where $\tilde{\text{Im}}$ stands for the imaginary part proportional to $\lambda^{(D-2)/2}$. Then
\begin{equation}
\tilde{\text{Im}} \Sigma_A^{(2)}(\vn{p},\lambda+\mathrm{i}
0^+)=-\frac{\pi}{2}\lambda^{(D-2)/2}\int \mathrm dq \int \mathrm d\Omega_D \,
V^2(q,p)\propto \lambda^{(D-2)/2}p^2
\end{equation}and
\begin{equation}
\tilde{\text{Im}} \Sigma_A^{(2)}(\vn{p},\lambda+\mathrm{i} 0^+)=-\tilde {\text{Im}} \Sigma_B^{(2)}(\vn{p},\lambda+\mathrm{i} 0^+)\:.
\end{equation} 
It follows that the amplitude ${\cal A}$ vanishes, because
the $\Sigma_C^{(2)}$ contribution is already of order $p^4$:
\begin{equation}
\tilde{\text{Im}} \Sigma_C^{(2)}(\vn{p},\lambda+\mathrm{i}
0^+)\propto -2 \lambda^{(D-2)/2}V(\vn{p},\vn{p})\Sigma^{(1)}(\vn p,z)\propto
  \lambda^{(D-2)/2}p^4.
\end{equation} 

In the following, we will show explicitily that the cancellation of
the $\lambda^{(D-2)/2}p^2$ term arises even for a given topological class
(quite large) of diagrams at $1/\rho^3$ order, and we will provide an
argument that predicts such cancellation at any pertubative order.

\section{The combinatorial computation}\label{SECT:COMBINATORIAL}

The first approach to the computation of the resolvent is
based on the expansion of Eq.~(\ref{RESOLVENT}) as a power series,
\begin{equation}
G(\vn{p}, z) = \sum_{R=1}^N
\frac{1}{z^{R+1}}\,\left(\lim_{N\to\infty} \overline{\frac{1}{N}\sum_{i,j=1}^N
  \mathrm{e}^{{\mathrm i} \vn{p}\cdot(\vn{x}_j -\vn{x}_i)}
  \left[\mat^R\right]_{ij}}\right)\,.
\label{EXPANSION}
\end{equation}
Although the final results will only depend on $p$, in order to
develop the formalism it is convenient to reintroduce the dependence on
$\vn{p}$.

\subsection{Organizing the calculation. The bare propagator.}

\subsubsection{Momentum shift: choosing wisely the integration order}
\label{MOMENT-SHIFT}

The $R$-th term of the expansion Eq.~(\ref{EXPANSION}) is
\begin{equation}
\sum_{i_0,i_1,\ldots i_R} \overline{\frac{1}{N}\sum_{i,j=1}^N
  \mathrm{e}^{{\mathrm i} \vn{p}\cdot(\vn{x}_{i_R} -\vn{x}_{i_0})}
  \left[\mat_{i_0,i_1} \mat_{i_1,i_2}\ldots \mat_{i_{R-1},i_R}\right]},
\label{EQ:TERM}
\end{equation}
where the average over the vibrational centers take the form of a
multi-dimensional integral with measure
$$\frac{1}{V^N}\int \prod_{i=1}^N \mathrm{d}\vn{x}_i.$$ As for all
such integrals, although the final result is independent of the order
in which the individual integrals are performed, the difficulties
encountered in a real computation are dramatically smaller if one
finds a wise ordering for iterated integrations.

Now consider the expression
\begin{equation}
\mathrm{e}^{-{\mathrm i} \vn{p} \cdot \vn{x}_{i_l}} \left[\delta_{i_l,i_{l+1}} \sum_{k_l\neq i_l} f(\vn{x}_{k_l}-\vn{x}_{i_l}) - [1-\delta_{i_l,i_{l+1}}] f(\vn{x}_{i_l} -\vn{x}_{i_{l+1}}) \right],
\end{equation}
which arises as a factor when introducing the explicit form
(Eq.~\ref{DEFERM}) of $\mat$ into Eq.~(\ref{EQ:TERM}). When dealing with
a diagonal term, we shall integrate over the position of the {\em
  medium} particle, $\vn{x}_{k_l}$; when dealing with an off-diagonal
term, we shall integrate over $\vn{x}_{i_l}$. For a diagonal term, the
integral over the position of the medium particle is easy, if the
particle index ${k_l}$ does not appear elsewhere in the chain (even if
the index $i_l$ is {\em sure} to appear at least once more along the
chain). For the non-diagonal term, the integral over $\vn{x}_{i_l}$ is
very simple if it does not appear later in the chain (even if
$i_{l+1}$ appears twice or more times in the chain, to the right). The
two integrals yield
$$ \frac{1}{V} [\hat f(\vn 0) \delta_{i_l,i_{l+1}} -
  [1-\delta_{i_l,i_{l+1}}]\hat f(\vn p)] \mathrm{e}^{-{\mathrm i} \vn{p}
  \cdot \vn{x}_{i_{l+1}}}.$$ Since a term of order $R$ has $R$ such
factors, the number of values the index $k_l$ (or $i_l$) can take
without violating the non-repetition condition is between $N$ and
$N-R$. But both $N/V$ and $(N-R)/V$ tend to $\rho$ in the
thermodynamic limit, hence momentum can shift through
non-index-repeating elements from left-to-right:
\begin{equation}
\mathrm{e}^{-{\mathrm i} \vn{p} \cdot \vn{x}_{i_l}} \left[\delta_{i_l,i_{l+1}} \sum_{k_l\neq i_l} f(\vn{x}_{k_l}-\vn{x}_{i_l}) - [1-\delta_{i_l,i_{l+1}}] f(\vn{x}_{i_l} -\vn{x}_{i_{l+1}}) \right]\longrightarrow 
\rho [\hat f(\vn 0) -\hat f(\vn p)] \mathrm{e}^{-{\mathrm i} \vn{p}
  \cdot \vn{x}_{i_{l+1}}}.
\end{equation}
Similarly, momentum can shift through non-repeating elements
from right-to-left. In that case, one would integrate over
$\vn{x}_{k_l}$ (diagonal term) or over $\vn{x}_{i_{l+1}}$ (non-diagonal):
\begin{equation}
\left[\delta_{i_l,i_{l+1}} \sum_{k_l\neq i_l} f(\vn{x}_{k_l}-\vn{x}_{i_l}) - [1-\delta_{i_l,i_{l+1}}] f(\vn{x}_{i_l} -\vn{x}_{i_{l+1}}) \right] \mathrm{e}^{{\mathrm i} \vn{p} \cdot \vn{x}_{i_{l+1}}}  \longrightarrow \mathrm{e}^{{\mathrm i} \vn{p}
  \cdot \vn{x}_{i_l}} 
\rho [\hat f(\vn 0) -\hat f(\vn p)] .
\end{equation}
Note that a given matrix-element might be considered as non-repeating
for momentum shift from right-to-left, but it could be not suitable
for the left-to-right momentum shift.

At this point, the computation of the leading order is straightforward. If
there are no obstacles for momentum shift, we just push to the 
left exponential $\mathrm{e}^{-{\mathrm i} \vn{p}\cdot
  \vn{x}_{i_0}}$ to the right until it cancels out with 
$\mathrm{e}^{{\mathrm i} \vn{p}\cdot \vn{x}_{i_R}}$, leaving us with
(since there are precisely $R$ matrix elements)
\begin{equation}
\rho^R[\hat f(\vn 0) -\hat f(\vn p)]^R.
\end{equation}
Then the high-density limit of the sum in Eq.~(\ref{EXPANSION}) is
\begin{equation}
G_0(\vn p,z)=\frac{1}{z-\epsilon(\vn p)}\,,
\end{equation}
which is then the bare propagator of the theory, as anounced in
Eq.~(\ref{DEF:G0}).

\subsubsection{Repeated indices}

Now consider a situation where we can shift the external
momentum $\vn{p}$ from left-to-right until a particular
particle-index (say $i_l=1$ or $k_l=1$) is repeated in the chain
somewhere to the right, so that we must stop. At this point, we 
shift the external momentum from right-to-left, until a particle-label
repetition $i_{r+1}=2$ or $k_r=2$ stop us. We depict this situation as
\begin{equation}
\ldots 1\, [\text{stuff}]\, 2\ldots.
\end{equation}
 At this point we will have
$$ \rho^S [\hat f(\vn 0)-\hat f(\vn p)]^S \mathrm{e}^{-{\mathrm i}\vn{p}\cdot
  \vn{x}_{i_l}}\,1\, [\text{stuff}]\, 2 \mathrm{e}^{{\mathrm
    i}\vn{p}\cdot \vn{x}_{i_{r+1}}}\, \rho^L [\hat f(\vn 0)-\hat f(\vn p)]^L.
$$ Since the very same scheme of particle-label repetitions
$1[\text{stuff}]2$ can be found for all values $L,S=0,1,2,\ldots$,
we can sum all those terms to find a contribution
$$ G_0(\vn{p},z) \mathrm{e}^{-{\mathrm i}\vn{p}\cdot \vn{x}_{i_l}} 1\,
[\text{stuff}]\, 2 \mathrm{e}^{{\mathrm i}\vn{p}\cdot \vn{x}_{i_{r+1}}}
G_0(\vn{p},z). $$
We interpret the two factors $G_0(\vn{p},z)$ as the external legs for
a Dyson resummation of the self-energy.

Clearly particle-label repetitions are going to be very
important in what follows, so some terminology will be useful. A
generic factor
\begin{equation}
\left[\delta_{i_l,i_{l+1}} \sum_{k_l\neq i_l} f(\vn{x}_{k_l}-\vn{x}_{i_l}) - [1-\delta_{i_l,i_{l+1}}] f(\vn{x}_{i_l} -\vn{x}_{i_{l+1}}) \right]\label{EC:GENERICO}
\end{equation}
will be called an {\em L-stop} if particles $k_l$ or $i_l$ are repeated
somewhere to the right (so that momentum cannot be shifted from
left-to-right trough index $i_l$). Similarly, we shall call it an {\em
  R-stop} if $k_l$ or $i_{l+1}$ are repeated somewhere to the
left. We note that a matrix element can be both a L-stop and a R-stop
(if $k_l$ is repeated both to the right and to the left, or if $i_l$
is repeated to the right while $i_{l+1}$ is repeated to the left). 

To make momentum flow through an L-stop or R-stops we resort to the so
called fake-particle trick. Consider a particle label, say $1$, that
appears twice (for instance, in an L-stop and in an R-stop to its
right). Before carrying out the average over $\{\vn{x}_i\}$, we
multiply the term by 1 written as
\begin{equation}
1=\int \mathrm{d}^D {\vn{y}}_{\tilde 1}\,\delta(\vn{x}_1 -\vn{y}_{\tilde
  1})=\frac{1}{(2\pi)^D} \int \mathrm{d}^D\vn{y}_{\tilde 1} \mathrm{d}^D
\vn{q}\, \mathrm{e}^{\mathrm{i} \vn{q} (\vn{x}_1-\vn{y}_{\tilde
    1})}.\label{EC:FASULLA1}
\end{equation}
Then we can pretend that particle $1$ takes two identities, $1$ and
$\tilde 1$, so that there is no repetition. The price we pay for this
simplification is that:
\begin{itemize}
\item we have an extra integration over the internal momentum $\vn{q}$,
\item we have to deal with an extra factor $\mathrm{e}^{\mathrm{i} \vn{q}
    \vn{x}_1}$ at the L-stop, and an extra $\mathrm{e}^{-\mathrm{i}
    \vn{q} \vn{y}_{\tilde 1}}$ at the R-stop, and
\item the fake particle $\vn{y}_{\tilde 1}$ does not bring a
  combinatorial $N$ factor, or an $1/V$ from the normalization of the
  $\vn{y}_1$ integral, so that there is a lacking factor of
  $\rho$ (this we can ignore if we add compensating $1/\rho$ to the
  final expression).
\end{itemize}
However, the modified momentum-shift formulae are simple enough to
justify these inconveniences. Integrating over $\vn{x}_1$ we obtain
\begin{subequations}
\begin{equation}
\mathrm{e}^{-{\mathrm i} \vn{p} \vn{x}_{i_l}} \mathrm{e}^{{\mathrm i}
  \vn{q} \vn{x}_{1}}
\left[\delta_{i_{l},i_{l+1}}f(\vn{x}_{1}-\vn{x}_{i_{l}})-(1-\delta_{1,i_{l+1}})\delta_{1,i_{l}}f(\vn{x}_{1}-\vn{x}_{i_{l+1}})\right]\longrightarrow
V(\vn{q},\vn{p}) \mathrm{e}^{-{\mathrm i} (\vn{p}-\vn{q})
  \vn x_{i_{l+1}}}.\label{EC:FASULLA2}
\end{equation}
Similarly, integrating over $\vn{y}_{\tilde 1}$ at the R-stop, we have
\begin{equation}
\left[\delta_{i_{r},i_{r+1}}f(\vn{y}_{\tilde 1}-\vn{x}_{i_{r}})-(1-\delta_{1,i_{r}})\delta_{\tilde
    1,i_{r+1}}f(\vn{x}_{i_{r}}-\vn{y}_{\tilde 1})\right] \mathrm{e}^{{\mathrm i} \vn{p}
  \vn{x}_{i_{r+1}}} \mathrm{e}^{-{\mathrm i} \vn{q} \vn{y}_{\tilde 1}}\longrightarrow
\mathrm{e}^{{\mathrm i} (\vn{p}-\vn{q})\vn{x}_{i_{r}}} V(\vn{q},\vn{p})\,.\label{EC:FASULLA3}
\end{equation}
\label{EC:FASULLA-GRP}
\end{subequations}

As a warning on momentum-shift, note that one may shift
momentum from left to right as long as {\em there is nothing to
the left still needing integration} (and similarly for right-to-left
shifts). Momentum shift can be visualized as a zip with two heads: one
pulls both heads until they meet (and then there are no integrals left
to be done).

\subsubsection{The reduction formula}\label{SECT:REDUCTION}

Imagine we face the situation
$$\ldots1[\text{stuff}]1\ldots,$$ i.e.\ the leftmost stop
is an L-stop at index position $l$, the rightmost stop is an
R-stop at index position $r+1$, and the particle that
prevents the two momentum shifts is the same at both ends, say
$i_l=i_{r+1}=1$ or any other possible combination ($k_l=i_{r+1}=1$,
$i_{l}=k_r=1$, or $k_l=k_{r}$). If the particle label $1$ does not
appear again inside the brackets, a nice reduction formula follows:
\begin{multline}
\mathrm{e}^{-{\mathrm i} \vn{p}
  \vn{x}_{i_l}}\left[\delta_{i_l,i_{l+1}} f( \vn{x}_{1}-
  \vn{x}_{i_{l}}) - [1-\delta_{1,i_{l+1}}]\delta_{i_l,1} f( \vn{x}_1 -
  \vn{x}_{i_{l+1}}) \right]
  \times [\text{stuff}]\times\\
  \left[\delta_{i_r,i_{r+1}} f( \vn{x}_{1}-
  \vn{x}_{i_r}) - [1-\delta_{i_r,1}]\delta_{i_{r+1},1} f( \vn{x}_{i_r}
  - \vn{x}_{1}) \right] \mathrm{e}^{{\mathrm i} \vn{p} 
  \vn{x}_{i_{r+1}}} \longrightarrow\\
 \frac{1}{\rho}\int\!\!\frac{\mathrm d^D\vn{q}}{(2\pi)^D} \,V^2( \vn{q},
      \vn{p}) \mathrm{e}^{-{\mathrm i} \vn{q} 
  \vn{x}_{i_{l+1}}} [\text{stuff}] \mathrm{e}^{+{\mathrm i} \vn{q} 
  \vn{x}_{i_{r}}}.
\label{EQ:REDUCCION}
\end{multline}
This can be proved by averaging over $ \vn{x}_1$. To adjust the power
of $\rho$, just recall that there were order $N$ choices for (say) the
index coincidence $k_l=i_{r+1}=1$. Note that the proof of
Eq.~(\ref{EQ:REDUCCION}) involves doing four different integrals. Let us
see how the fake-particle formulae (Eqs.~(\ref{EC:FASULLA-GRP})) yield
the same result effortlessly. The introduction of the fake particle
transforms the left-hand side of Eq.~(\ref{EQ:REDUCCION}) to
\begin{multline}
\mathrm{e}^{-{\mathrm i} \vn p \vn{x}_{i_l}} \mathrm{e}^{{\mathrm i} \vn q
  \vn{x}_{1}}
\left[\delta_{i_{l},i_{l+1}}f(\vn{x}_{1}-\vn{x}_{i_{l}})-(1-\delta_{1,i_{l+1}})\delta_{1,i_{l}}f(\vn{x}_{1}-\vn{x}_{i_{l+1}})\right]
\times[\text{stuff}]\times\\
   \left[\delta_{i_{r},i_{r+1}}f(\vn{y}_{\tilde
    1}-\vn{x}_{i_{r}})-(1-\delta_{1,i_{r}})\delta_{\tilde
    1,i_{r+1}}f(\vn{x}_{i_{r}}-\vn{y}_{\tilde 1})\right]
\mathrm{e}^{{\mathrm i} \vn{p} \vn{x}_{i_{r+1}}} \mathrm{e}^{-{\mathrm i} \vn{q}
  \vn{y}_{\tilde 1}}
\end{multline}
We now merely shift momentum to the right using Eq.~(\ref{EC:FASULLA2})
and to the left using Eq.~(\ref{EC:FASULLA3}) to obtain
$$\frac{1}{\rho}\int\!\!\frac{\mathrm d^D \vn{q}}{(2\pi)^D} \,V( \vn{q},
\vn{p})^2 \, \mathrm{e}^{-{\mathrm i} (\vn{p}-\vn{q}) 
  \vn{x}_{i_{l+1}}}\ \times[\text{stuff}]\times\ \mathrm{e}^{+{\mathrm i} (\vn{p}-\vn{q})
  \vn{x}_{i_{r}}}.$$ Now a change of integration variable
$\vn{q}\longrightarrow \vn{p}-\vn{q}$ and the second of identities
(\ref{IDENTITIES}) yield Eq.~(\ref{EQ:REDUCCION}).

Both $\Sigma^{(1)}$ and $\Sigma^{(2)}_A$ follow directly from
Eq.~(\ref{EQ:REDUCCION}). Also, more general expresions can be
found easily from it, as we shall see below.

\subsection{Order $1/\rho$}
If no further particle-label repetition arise, the momentum
$\mathrm{e}^{-{\mathrm i} \vn{q} \cdot \vn{x}_{i_{l+1}}}$ in
Eq.~(\ref{EQ:REDUCCION}) can be shifted to the right until it is
killed by the second exponential $\mathrm{e}^{{\mathrm i} \vn{q} \cdot
  \vn{x}_{i_{r}}}$. We have then a set of contributions of the form
\begin{equation}
\sum_{a+b+c+2=R}
\frac{[\rho (\hat f(\vn 0)-\hat f(\vn p))]^a}{z^{a+1}}\times
\frac{1}{\rho}\int\!\!\frac{\mathrm d^D\vn{q}}{(2\pi)^D} V^2(\vn{q},\vn{p}) 
 \frac{[\rho (\hat f(\vn 0)-\hat f(\vn q))]^b}{z^{b+1}}\times
 \frac{[\rho (\hat f(\vn 0)-\hat f(\vn p))]^c}{z^{c+1}} ,\label{INTERMEDIA}
\end{equation}
composed of the product of three harmonic series that are easily seen
to add-up to
\begin{equation}
G_0(\vn{p},z)\times \frac{1}{\rho}\int\!\!\frac{\mathrm d^D\vn{q}}{(2\pi)^D}
   V^2(\vn{q},\vn{p}) G_0(\vn{q},z)\times G_0(\vn{p},z).
\end{equation}
If we interpret the two factors $G_0(\vn{p},z)$ as external legs of a Dyson
resummation, we get
\begin{equation}
\Sigma^{(1)}(\vn{p},z)=
\frac{1}{\rho}\int\!\!\frac{\mathrm d^D\vn{q}}{(2\pi)^D}
 V^2(\vn{q},\vn{p}) G_0(\vn{q},z),\label{EQ:SIGMA1}
\end{equation}
which is the first order result anticipated in sec.~\ref{SECT:MAIN}.

\subsection{Order $1/\rho^2$}
\label{SECT:second-order}

The cases with two pairs of repeated indices, or one index occurring
three times contribute to the second-order corrections. The
contributions separate naturally in three kinds, according to the
arrangement of the repeated indices.

\subsubsection{The nested case: $\Sigma^{(2)}_A$}

\label{SECT:CACTUS}
Take now the scheme of particle repetitions giving rise to
Eq.~(\ref{INTERMEDIA}), and place it {\em in between} an external
pair of particle repetitions:
$$\ldots 2\ldots 1\ldots 1\ldots 2\ldots. $$ Assume that the index $2$
happens twice and only twice in the chain. We are
thus entitled to use the reduction formula, Eq.~(\ref{EQ:REDUCCION}), for
particle 2. The inner momentum $\vn{q}$, can then be shifted (from
either side) until it hits particle 1, where it produces a
contribution such as Eq.~(\ref{INTERMEDIA}). The only difference is in
that the role previously played by the external momentum $\vn{p}$ is
now played by the internal momentum $\vn{q}$. We get, without need for
further computation,
\begin{equation}
\Sigma_A^{(2)}(\vn{p},z) =\frac{1}{\rho}\int\!\!\frac{\mathrm d^D
  \vn{q}}{(2\pi)^3} V(\vn{q},\vn{p})
G_0(\vn{q},z)\Sigma^{(1)}(\vn{q},z) G_0(\vn{q},z) V(\vn{q},\vn{p}). 
\end{equation}

\subsubsection{The interleaved case: $\Sigma^{(2)}_B$}
\label{SECT:SIGMA_B}

The $\Sigma_B^{(2)}$ piece in Eq.~(\ref{EQ:SIGMA2-TRES-PARTES}) arises from
the pattern $$\ldots 1\ldots 2\ldots 1\ldots2\ldots.$$ A moment's thought
indicates that 
the leftmost $1$ must belong to an L-stop, while the rightmost $2$
must be an R-stop. Furthermore, the internal 2 should be an L-stop
(otherwise, one would use a fake particle to shift momentum from
left-to right over it trivially). For the same reason, the
internal 1 should belong to an R-stop.

Our previous succes with the reduction formula, Eq.~(\ref{EQ:REDUCCION}) suggests
that we try to deal with all such terms at once, by performing the integral
\begin{multline}
\int\!\!\frac{\mathrm d^{D}\vn{x}_{1}\mathrm d^{D}\vn{x}_{2}}{V^{2}}\, {\mathrm e}^{-{\mathrm
    i} \vn{p}\vn{x}_{i_l}}\left[\delta_{i_{l},i_{l+1}}
  f(\vn{x}_{1}-\vn{x}_{i_{l}})-(1-\delta_{1,i_{l+1}})
  \delta_{1,i_{l}}f(\vn{x}_{1}-\vn{x}_{i_{l+1}})\right] \left[\ldots\right]\times\\
  \left[\delta_{i_{r},i_{r+1}}f(\vn{x}_{2}-\vn{x}_{i_{r}}) -
    (1-\delta_{2,i_{r+1}})\delta_{2,i_{r}}f(\vn{x}_{2}-\vn{x}_{i_{r+1}})\right]
  \left[\ldots\right]\times\\
 \left[\delta_{i_{s},i_{s+1}}f(\vn{x}_{1}-\vn{x}_{i_{s}}) -
   (1-\delta_{1,i_{s}})\delta_{1,i_{s+1}}
   f(\vn{x}_{i_{s}}-\vn{x}_{1})\right] \left[\ldots\right]\times\\
   \left[\delta_{i_{z},i_{z+1}}f(\vn{x}_{2}-\vn{x}_{i_{z}}) -
     (1-\delta_{2,i_{z}})\delta_{2,i_{z+1}} f(\vn{x}_{i_{z}}-\vn{x}_{2})\right]
   {\mathrm e}^{{\mathrm i} \vn{p} \vn{x}_{i_{z+1}}}.\label{EC:B4}
\end{multline}
Here the several $[\ldots]$ stand for arbitrary numbers of matrix
elements without momentum stops arises. Note that not all the $\ldots
1\ldots2\ldots1\ldots2\ldots$ terms have the form Eq.~(\ref{EC:B4}): 
the central $2,1$ particles may also collapse onto a single matrix
element (necessarily non-diagonal) which is both an R-stop and an
L-stop. Such terms will be considered in sec.~\ref{SECT:SIGMA_C}.

We now introduce two fake particles, $\tilde 1$ and $\tilde 2$, to
transform the above integral into
\begin{multline}
  \int\!\!\frac{\mathrm d^{D}\vn{q}
    \mathrm d^{D}\vn{k}}{(2\pi)^{2D}}\int\!\!\frac{\mathrm d^{D}\vn{x}_{1}\mathrm d^{D}\vn{y}_{\tilde
      1}\mathrm d^{D}\vn{x}_{2}\mathrm d^{D}\vn{y}_{\tilde 2}}{V^{2}}\,  {\mathrm
    e}^{-{\mathrm i}\vn{p} \vn{x}_{i_l}}
  \left[\delta_{i_{l},i_{l+1}}e^{\mathrm{i} \vn{q}
      \vn{x}_{1}}f(\vn{x}_{1}-\vn{x}_{i_{l}})-(1-\delta_{1,i_{l+1}})\delta_{1,i_{l}}e^{iq\vn{x}_{1}}f(\vn{x}_{1}-\vn{x}_{i_{l+1}})\right]
  \left[\ldots\right]\times\\
   \left[\delta_{i_{r},i_{r+1}}e^{ik\vn{x}_{2}}f(\vn{x}_{2}-\vn{x}_{i_{r}})-(1-\delta_{2,i_{r+1}})\delta_{2,i_{r}}e^{ik\vn{x}_{2}}f(\vn{x}_{2}-\vn{x}_{i_{r+1}})\right]
  \left[\ldots\right]\times\\
   \left[\delta_{i_{s},i_{s+1}}e^{-\mathrm{i} \vn{q} \vn{y}_{1}}f(\vn{y}_{\tilde
      1}-\vn{x}_{i_{s}})-(1-\delta_{\tilde{1},i_{s}})\delta_{\tilde{1},i_{s+1}}e^{-\mathrm{i}
      \vn{q} \vn{y}_{1}}f(\vn{x}_{i_{s}}-\vn{y}_{\tilde 1})\right]
  \left[\ldots\right]\times \\
   \left[\delta_{i_{z},i_{z+1}}e^{-\mathrm{i}   \vn{k} \vn{y}_{2}}f(\vn{y}_{\tilde
      2}-\vn{x}_{i_{z}})-(1-\delta_{\tilde{2},i_{z}})\delta_{\tilde{2},i_{z+1}}e^{-i
      \vn{k}\vn{y}_{\tilde 2}}f(\vn{x}_{i_{z}}-\vn{y}_{\tilde
      2})\right]{\mathrm e}^{{\mathrm i}\vn{p} \vn{x}_{i_{z+1}}}.
\end{multline}
One then shifts momentum from left-to-right up to $i_s$ and from
right-to-left again up to $i_s$, to find
\begin{equation}
\frac{1}{\rho^{2}}\int\!\!\frac{\mathrm d^D\vn{q} \mathrm d^D\vn{k}}{(2\pi)^{2D}}\,
 V(\vn{q},\vn{p})G_{0}(\vn{p}-\vn{q},z) V(\vn{k},\vn{p}-\vn{q})
 G_{0}(\vn{p}-\vn{q}-\vn{k},z)V(\vn{q},\vn{p}-\vn{k})
 G_{0}(\vn{p}-\vn{k},z)V(\vn{k},\vn{p}).
\end{equation} 
Eq.~(\ref{EC:SIGMAB}) follows after changing integration variables
according to
\begin{equation}
\vn{p}-\vn{q}\to \vn{q}, \qquad \text{and} \qquad
\vn{q}-\vn{k}\to \vn{k}.
\end{equation}

\subsubsection{The collapse of a L-stop and a R-stop: $\Sigma_C^{(2)}$}
\label{SECT:SIGMA_C}
As we have remarked, it can happen that the L-stop and R-stop of
Eq.~(\ref{EC:B4}) actually belong to the same matrix element,
necessarily non-diagonal. However, any non-diagonal term should be
paired with a diagonal one.  As we mentioned in sec.~\ref{MOMENT-SHIFT}, a
diagonal term can be both a L-stop and a R-stop if the medium particle
is repeated both to the left and to the right. Hence we will be
considering here this kind of terms ($O$:of diagonal matrix element,
$D$: diagonal matrix element):
\begin{equation}
1\ldots O(21)\ldots 2\quad +\quad 1\ldots D(1)\ldots 1
\end{equation}
The the terms with an off-diagonal index appearing three times
($1\ldots O(1?)\ldots 1$) do not belong to $\Sigma_C^{(2)}$, and are
considered in sec.~\ref{SECT:DYSON-ORDEN-2}.

Let us start with the case $1\ldots D(1)\ldots 1$:
\begin{multline}
  \int\!\!\frac{\mathrm d^{D}\vn{x}_{1}}{V} {\mathrm e}^{-{\mathrm
      i}\vn{p}\vn{x}_{i_{l}}} \left[\delta_{i_{l},i_{l+1}}
    f(\vn{x}_{1}-\vn{x}_{i_{l}}) - (1-\delta_{1,i_{l+1}})
    \delta_{1,i_{l}} f(\vn{x}_{1}-\vn{x}_{i_{l+1}}) \right]
  \left[\ldots\right] \times\\
   \delta_{i_{r},i_{r+1}} f(\vn{x}_{1}-\vn{x}_{i_{r}}) \left[\ldots\right]\times\\
   \left[\delta_{i_{s},i_{s+1}} f(\vn{x}_{1}-\vn{x}_{i_{s}}) -
     (1-\delta_{1,i_{s}}) \delta_{1,i_{s+1}}
     f(\vn{x}_{i_{s}}-\vn{x}_{1}) \right] {\mathrm e}^{{\mathrm
       i}\vn{p}\vn{x}_{i_{s+1}}}.
\end{multline}
We now introduce two extra fake particles to substitute particle $1$,
namely $\tilde 1$ and $\hat 1$ via the identity
\begin{equation}
1=\int\!\!\mathrm d^D\vn{y}_{\tilde 1} \int\!\!\mathrm d^D\vn{z}_{\hat
  1}\, \delta(\vn{x}_{1}-\vn{y}_{\tilde 1}) \delta(\vn{y}_{\tilde
  1}-\vn{z}_{\hat 1}) = \frac{1}{(2\pi)^{2D}}\int\!\!\mathrm d^D\vn{q}\,
  \mathrm{d}^D \vn{k} \, \mathrm d^D\vn{y}_{\tilde 1}\, \mathrm d^D\vn{z}_{\hat 1}\,
  {\mathrm e}^{{\mathrm i} \vn{q}
  (\vn{x}_{1}-\vn{y}_{\tilde 1})} {\mathrm e}^{{\mathrm i} \vn{k}
  (\vn{y}_{\tilde 1}-\vn{z}_{\hat 1})},
\end{equation}
to find
\begin{multline}
\int\!\!\frac{\mathrm d^{D} \vn{q} \mathrm d^{D}\vn{k}}{(2\pi)^{2D}}
\int\!\!\frac{\mathrm d^{D}\vn{x}_{1}{\mathrm d^{D}\vn{y}_{\tilde 1}} \mathrm d^{D}\vn{z}_{\hat
      1}}{V} \,{\mathrm e}^{- {\mathrm
    i}\vn{p}\vn{x}_{i_{l}}}\left[\delta_{i_{l},i_{l+1}}f(\vn{x}_{1}-\vn{x}_{i_{l}})-(1-\delta_{1,i_{l+1}})\delta_{1,i_{l}}f(\vn{x}_{1}-\vn{x}_{i_{l+1}})\right]{\mathrm
  e}^{{\mathrm i} \vn{q}\vn{x}_{1}}\left[\ldots\right]\times\\
  \delta_{i_{r},i_{r+1}}f(\vn{y}_{\tilde
  1}-\vn{x}_{i_{r}}){\mathrm e}^{{\mathrm
    i}(\vn{k}-\vn{q})\tilde{\vn{x}_{1}}} \left[\ldots\right]
  {\mathrm e}^{-{\mathrm i} \vn{k}\vn{z}_{\hat
      1}}\left[\delta_{i_{s},i_{s+1}}f(\vn{z}_{\hat
      1}-\vn{x}_{i_{s}})-(1-\delta_{\hat{1},i_{s+1}})\delta_{\hat{1},i_{s+1}}f(\vn{x}_{i_{s}}-\vn{z}_{\hat
      1})\right]{\mathrm e}^{-{\mathrm i}\vn{p}\vn{x}_{i_{s+1}}}\:.
\end{multline}
Finally we shift momentum from left-to right up to $i_r$, from
right-to-left up to $i_{r+1}$ and integrate over $\tilde
{\vn x}_1$. We obtain
\begin{equation} 1\ldots D(1)\ldots1=
\frac{1}{\rho^{2}}\int\!\!\frac{d^{D}\vn{q}d^{D}\vn{k}}{(2\pi)^{2D}}
V(\vn{q},\vn{p})G_{0}(\vn{p}-\vn{q},z)
     [\rho \hat f(\vn{k}-\vn{q})]
     G_{0}(\vn{p}-\vn{k})V(\vn{k},\vn{p}). \label{EC:SIGMAC-D}
\end{equation}

Consider now $1\ldots O(21)\ldots 2$:
\begin{multline}
  \int\!\!\frac{\mathrm d^{D}\vn{x}_{1} \mathrm d^{D}\vn{x}_{2}}{V^2}
           {\mathrm e}^{-{\mathrm i}
             \vn{p}\vn{x}_{i_{l}}}\left[\delta_{i_{l},i_{l+1}}
             f(\vn{x}_{1}-\vn{x}_{i_{l}})-(1-\delta_{1,i_{l+1}})
             \delta_{1,i_{l}}f(\vn{x}_{1}-\vn{x}_{i_{l+1}})\right]
           \left[\ldots\right]\times \\
           [-\delta_{i_{r},2}] \delta_{i_{r+1},1}
           f(\vn{x}_{2}-\vn{x}_{1}) \left[\ldots\right]
           \left[\delta_{i_{s},i_{s+1}}f(\vn{x}_{2}-\vn{x}_{i_{s}}) -
             (1-\delta_{2,i_{s}}) \delta_{2,i_{s+1}}
             f(\vn{x}_{i_{s}}-\vn{x}_{2}) \right]{\mathrm
             e}^{{\mathrm i}\vn{p}\vn{x}_{i_{s+1}}}.
\end{multline}
Introducing two fake particles, $\tilde 1$ and $\tilde 2$ we can rewrite it as
\begin{multline}
  \int\!\!\frac{\mathrm d^{D}\vn{q} \mathrm d^{D}\vn{k} \mathrm d^{D} \vn{x}_{1}
    \mathrm d^{D}\vn{x}_{2}} {(2\pi)^{2D}V^2}
    {\mathrm e}^{-{\mathrm i}\vn{p}\vn{x}_{i_{l}}} {\mathrm e}^{{\mathrm
      i}\vn{q}\vn{x}_1}
  \left[\delta_{i_{l},i_{l+1}}f(\vn{x}_{1}-\vn{x}_{i_{l}}) -
    (1-\delta_{1,i_{l+1}}) \delta_{1,i_{l}}
    f(\vn{x}_{1}-\vn{x}_{i_{l+1}})\right] \left[\ldots\right]\times\\
   \left[-{\mathrm e}^{-{\mathrm i} \vn{q} \vn{y}_{\tilde 1}}\right] 
   \delta_{i_{r},2}\delta_{i_{r+1},\tilde 1}
   f(\vn{x}_{2}-\vn{y}_{\tilde 1}) {\mathrm e}^{{\mathrm i}\vn{k} 
    \vn{x}_2} \left[\ldots\right] \left[\delta_{i_{s},i_{s+1}}
  f(\vn{x}_{2}-\vn{x}_{i_{s}})- (1-\delta_{\tilde 2,i_{s}})
  \delta_{\tilde 2,i_{s+1}} f(\vn{x}_{i_{s}}-\vn{y}_{\tilde 2})\right]
  {\mathrm e}^{-{\mathrm i} \vn{k} \vn{y}_{\tilde 2}} {\mathrm
    e}^{{\mathrm i}\vn{p} \vn{x}_{i_{s+1}}}.
\end{multline}
Shifting momentum left-to-right up to  $i_r$ and from right-to-left up
to $i_{r+1}$ we are left with
\begin{equation}
-\frac{1}{V} \int\!\! \mathrm d^D\vn{x}_2 \mathrm d^D\vn{y}_{\tilde 1}\,
  f(\vn{x}_{2}-\vn{y}_{\tilde 1}) {\mathrm e} ^{-{\mathrm i}
  (\vn{p}-\vn{q}-\vn{k}) (\vn{x}_2-\vn{x}_1)}= -\hat f(\vn{p}-\vn{q}-\vn{k}),
\end{equation}
so that adjusting the power of $\rho$, we get
\begin{equation}
1\ldots O(21)\ldots 2=
\frac{1}{\rho^{2}}\int\!\!\frac{\mathrm d^{D}\vn{q}\mathrm d^{D}\vn{k}}{(2\pi)^{2D}}
\,V(\vn{q},\vn{p}) G_{0}(\vn{p}-\vn{q},z)
     [-\rho \hat
     f(\vn{p}-\vn{q}-\vn{k})]G_{0}(\vn{p}-\vn{k},z)V(\vn{k},\vn{p}).
\label{EC:SIGMAC-O}
\end{equation}
Adding together the two pieces, Eqs.~(\ref{EC:SIGMAC-O})
and~(\ref{EC:SIGMAC-D}) we finally find
\begin{equation}
\Sigma_C^{(2)}=\frac{1}{\rho^{2}}
\int\!\!\frac{\mathrm d^{D}\vn{q}\mathrm d^{D}\vn{k}}{(2\pi)^{2D}} V(\vn{q},\vn{p})
G_{0}(\vn{p}-\vn{q},z) V(\vn{k}-\vn{q},\vn{p}-2\vn{q})
G_{0}(\vn{p}-\vn{k},z)V(\vn{k},\vn{p}),
\end{equation}
which after the change of variables
\begin{equation}
\vn{q}\longrightarrow \vn{p}-\vn{q},\qquad \vn{k}\longrightarrow \vn{p}-\vn{k},
\end{equation}
and use of identities Eq.~(\ref{IDENTITIES}) yield Eq.~(\ref{EQ:SIGMA_C}).

\subsubsection{The Dyson resummtion to order $1/\rho^2$}\label{SECT:DYSON-ORDEN-2}

Recalling Eq.~(\ref{EC:EXPANSION}), we notice that we have still not
identified the pattern of particle-label repetitions that gives
rise to the second-order terms appearing in the Dyson resummation the
first-order self-energy,
\begin{equation}
\Sigma^{(1)}(\vn{p},z)G_0(\vn{p},z)\Sigma^{(1)}(\vn{p},z),
\label{EC:DYSON-ORDEN-2}
\end{equation}
(we have not written the irrelevant external legs). The natural candidate is
\begin{equation}
1\ldots1\ldots 2\ldots 2,
\end{equation}
where the sequence is L-stop, R-stop, L-stop, R-stop. This expectation
is correct, but it will turn out that the constraint imposed by the
matrix-product structure needs extra terms to build
Eq.~(\ref{EC:DYSON-ORDEN-2}). These missing terms will be provided by the
pattern $1\ldots O(1?)\ldots 1$.

Let us first compute blindly the term $1\ldots1\ldots 2\ldots 2$,
incurring in a quite instructive mistake. We introduce only one
fake particle, $\tilde 1$:
\begin{multline}
{\mathrm e}^{-{\mathrm i}\vn{p} \vn{x}_{i_{l}}} {\mathrm e}^{{\mathrm
    i} \vn{q} \vn{x}_1} \left[\delta_{i_{l},i_{l+1}}
  f(\vn{x}_{1}-\vn{x}_{i_{l}}) - (1-\delta_{1,i_{l+1}})
  \delta_{1,i_{l}} f(\vn{x}_{1}-\vn{x}_{i_{l+1}})\right]
\left[\ldots\right]\times\\
  {\mathrm e}^{-{\mathrm i} \vn{q} \vn{y}_{\tilde 1}}
  \left[\delta_{i_{r},i_{r+1}} f(\vn{y}_{\tilde 1}- \vn{x}_{i_{r}})-
    (1-\delta_{\tilde 1,i_{r}}) \delta_{\tilde 1,i_{r+1}}
    f(\vn{x}_{i_{r}}-\vn{y}_{\tilde 1}) \right]
  \left[\ldots\right]\times\\
   \left[\delta_{i_{s},i_{s+1}} f(\vn{x}_{2}-\vn{x}_{i_{s}})-
     (1-\delta_{2,i_{s+1}}) \delta_{2,i_{s}}
     f(\vn{x}_{2}-\vn{x}_{i_{s+1}})\right] \left[\ldots\right]\times\\
    \left[\delta_{i_{z},i_{z+1}} f(\vn{x}_2-\vn{x}_{i_{r}})-
      (1-\delta_{2,i_{z}})\delta_{2,i_{z+1}} f(\vn{x}_{i_{r}}-\vn{x}_2)\right]
     {\mathrm e}^{{\mathrm i} \vn{p} \vn{x}_{i_{z+1}}}.
\label{EC:PROBLEMA}
\end{multline}
We shift momentum from left to right up to  $i_r$ as usual. At this
point, we still need to push the momentum to the right ({\em this} is
unusual). We  need to perform two integrals,
\begin{align}
\int \!\!\mathrm d^D\vn{y}_{\tilde 1} \, {\mathrm e}^{-{\mathrm
    i}(\vn{p}-\vn{q})\vn{x}_{i_{r}}} {\mathrm e}^{-{\mathrm
    i}\vn{q}\vn{y}_{\tilde
    1}} \left[\delta_{i_{r},i_{r+1}}f(\vn{y}_{\tilde
    1}-\vn{x}_{i_{r}})\right]&= \hat f(\vn q){\mathrm e}^{{\mathrm
    i} \vn{p} \vn{x}_{i_{s+1}}},  \\
    \int \!\!\mathrm d^D\vn{x}_{i_r} \, {\mathrm
  e}^{-{\mathrm i}(\vn{p}-\vn{q})\vn{x}_{i_{r}}} {\mathrm
  e}^{-{\mathrm i}\vn{q}\vn{y}_{\tilde 1}} \left[-(1-\delta_{\tilde
  1,i_{r}})\delta_{\tilde  1,i_{r+1}}f(\vn{x}_{i_{r}}-\vn{y}_{\tilde
  1})\right]&=
   -\hat f(\vn{p}-\vn{q}){\mathrm
  e}^{{\mathrm i} \vn{p} \vn{x}_{i_{s+1}}}.
\end{align}
Hence the integrations up to this point yield
\begin{equation}
\frac{1}{\rho}\int\!\!\frac{\mathrm d^D \vn{q}}{(2\pi)^D}\,
  V(\vn{q},\vn{p})G_0(\vn{p}-\vn{q},z)V(\vn{q},\vn{p}) {\mathrm
  e}^{{\mathrm i}\vn{p} \vn{x}_{i_{s+1}}}=\Sigma^{(1)}(\vn{p},z) {\mathrm e}^{{\mathrm
    i}\vn{p} \vn{x}_{i_{s+1}}}.
\end{equation}
It seems to be an easy matter to complete the computation: one
pushes momentum $p$ to the right up to $i_s$, seemingly yielding
a bare propagator $G_0(\vn p,z)$, and we would be left with $2\ldots2$ (a
standard diagram for the self-energy at order $1/\rho$). However,
after some reflection it is clear that an R-stop and an L-stop such as
$\ldots 1\ldots 2\ldots$, where
{\em both particle $1$ and particle $2$ appear in off-diagonal matrix
  elements} must be separated by {\em at least} one off-diagonal
matrix element. Hence if there are $S$ matrix elements
between the R-stop and the L-stop, when shifting momentum $p$ we will
encounter a factor
$$\rho^S [\hat f(\vn 0)-\hat f(\vn p)]^S - [\rho \hat f(\vn 0)]^S$$
which adding the geometric series means
$$G_0(\vn p,z)-\frac{1}{z-\rho \hat f(\vn 0)}.$$ Hence the correct
result is
\begin{multline}
1\ldots1\ldots2\ldots2 =
\Sigma^{(1)}(\vn{p},z)G_0(\vn{p},z)\Sigma^{(1)}(\vn{p},z)-\\
 -\frac{1}{z-\rho\hat f(\vn 0)} \frac {1}{\rho^2}
 \int\!\!\frac{\mathrm d^D\vn{q}\mathrm d^D\vn{k}}{(2\pi)^{2D}}\,
 V(\vn{q},\vn{p})G_0(\vn{p}-\vn{q},z) [\rho\hat f(\vn{p}-\vn{q})]
 [\rho\hat f(\vn{p}-\vn{k})] G_0(\vn{p}-\vn{k},z)V(\vn{k},\vn{p}).
\label{EC:FASTIDIO}
\end{multline}

We will now show that the second term in Eq.~(\ref{EC:FASTIDIO}) is
cancelled by the contribution from
$$1\ldots O(1?)\ldots 1.$$
In this pattern,  the leftmost 1 belongs to an L-stop and the
rightmost one to an R-stop.  The first observation is that the central 1
in the $O(1?)$ must nessarily appear in an R-stop (because we never
find the same particle in any matrix element $f(\vn{x}_i-\vn{x}_j)$, and due to
the constraint imposed by the matrix product). The second observation
is that there must be, at least, one off-diagonal matrix element between
the two R-stops sharing the common particle 1.
Introducing fake particles $1$ and $\tilde 1$, we are left with
\begin{multline}
\int\!\!\frac{\mathrm d^{D}\vn{q}\mathrm d^{D}\vn{k}}{(2\pi)^{2D}}
\int\!\!\frac{\mathrm d^{D}\vn{x}_{1}\mathrm d^{D}\vn{y}_{\tilde 1}\mathrm d^{D}\vn{z}_{\hat
    1}}{V} \,
\mathrm{e}^{-\mathrm{i}\vn{p}\vn{x}_{i_{l}}}\left[\delta_{i_{l},i_{l+1}}f(\vn{x}_{1}-\vn{x}_{i_{l}})-
  (1-\delta_{1,i_{l+1}})\delta_{1,i_{l}}
  f(\vn{x}_{1}-\vn{x}_{i_{l+1}})\right]
\mathrm{e}^{\mathrm{i}\vn{q}\vn{x}_{1}} \left[\ldots\right]\times\\
     \left[-(1-\delta_{\tilde{1},i_{r+1}})
       \delta_{\tilde{1},i_{r+1}}f(\vn{x}_{i_{r}}-\vn{y}_{\tilde
         1})\right] \mathrm{e}^{\mathrm{i}(\vn{k}-\vn{q})
       \vn{y}_{\tilde 1}} \left[\ldots\right]'\times \\
    \mathrm{e}^{-\mathrm{i}\vn{k}\vn{z}_{\hat
    1}}\left[\delta_{i_{s},i_{s+1}}f(\vn{z}_{\hat
    1}-\vn{x}_{i_{s}})-(1-\delta_{\hat{1},i_{s}})
  \delta_{\hat{1},i_{s+1}}f(\vn{x}_{i_{s}}-\vn{z}_{\hat 1})\right]
\mathrm{e}^{-\mathrm{i}\vn{p}\vn{x}_{i_{s+1}}}.
\label{EC:LIGADURA}
\end{multline}
All that remains is a simple momentum shift, keeping in mind that when
going over the factor $[\ldots]'$ it will give
\begin{equation}
G_0(\vn{p}-\vn{k},z)-\frac{1}{z-\rho\hat f(\vn 0)}=-\rho\hat
f(\vn{p}-\vn{k}) \frac{G_0(\vn{p}-\vn{k},z)}{z-\rho\hat f(\vn 0)}.
\end{equation}
Thus one finally finds
\begin{equation}
1\ldots O(1?)\ldots 1 = \frac{1}{z-\rho\hat f(\vn 0)} \frac {1}{\rho^2}
\int\!\!\frac{\mathrm d^D\vn{q} \mathrm d^D\vn{k}}{(2\pi)^{2D}} \,
 V(\vn{q},\vn{p})G_0(\vn{p}-\vn{q},z) [\rho\hat
  f(\vn{p}-\vn{q})][[\rho\hat
    f(\vn{p}-\vn{k})]G_0(\vn{p}-\vn{k},z)V(\vn{k},\vn{p}).
\end{equation}

\subsection{Higher orders}

The argument of sec.~\ref{SECT:CACTUS} is fully general. Consider
the contribution of order $1/\rho^n$ to the {\em propagator,} rather
than the self-energy (i.e.\ let us include both the connected and
disconnected pieces). We can write this as
$G_0(\vn{p},z)H^{(n)}(\vn{p},z) G_0(\vn{p},z)$. Let
us emphasize that $H^{(n)}(\vn{p},z)$ refers to the full contribution
to the propagator at order $1/\rho^n$, not to a particular topological
subset (such as the cactus~\cite{ERM3}).

We may enclose the scheme of particle label repetitions that generates
$H^{(n)}(\vn{p},z)$ within an L-stop and an R-stop with equal particle
labels that do not appear again along the chain. Under such
circumstances, we are entitled to use the reduction formula,
Eq.~(\ref{EQ:REDUCCION}), which yields
\begin{equation}
\Sigma_A^{(n+1)}(\vn{p},z) =
\frac{1}{\rho}\int\frac{\mathrm d^D \vn{q}}{(2\pi)^D}\, V(\vn{q},\vn{p}) G_0(\vn{q},z) H^{(n)}(\vn{q},z) G_0(\vn{q},z) V(\vn{q},\vn{p})\,.\label{EC:q2-GENERICA}
\end{equation}
Clearly this is not the full self-energy at order
$1/\rho^{n+1}$, but it is a genuine part of it that automatically
verifies
\begin{equation}
\Sigma_A^{(n+1)}(\vn{p},z) = 0.
\end{equation}
In particular if $n=1$ this gives the $1/\rho^2$ term
$\Sigma_A^2(\vn{p},z)$ discussed above. It is interesting to note that
\begin{equation}
\text{Im}\Sigma_A^{(n+1)}(\vn{p},\lambda+\mathrm{i}0^+)\sim
p^2\lambda^{(D-2)/2}  ,
\end{equation}
since for $q\sim 1$, and $z=\lambda+\mathrm{i}0^+$, for small
$\lambda$ it is expected that (Debye spectrum, see Sect.~\ref{SECT:EXAMPLE})
\begin{equation}
\text{Im} H^{(n)}(\vn{q},z)\propto \lambda^{(D-2)/2}.
\end{equation}
Thus, the vanishing of the amplitude ${\cal A}$ in Eq. (\ref{WRONG})
implies that non trivial cancellations occur at \emph{all orders} in
perturbation theory. Since we are presenting arguments for such
cancellation, we agree with Ref.~\cite{SCHIRMACHER} in that the
$\Sigma_A^{(n+1)}$ terms alone do not reproduce the correct analytic
structure of the theory.

\subsubsection{Towards the self-energy at third order}

Using the combinatorial rules described above, it is possible
to push the perturbative computation to order $1/\rho^3$ order, which has
never been attempted before. Here we will limit ourselves to the
terms without collapse of an $R$-stop with an $L$-stop (i.e.\ we
will retain only the terms with 6 vertex functions). The reason
is that the combinatorial computation suggests very simple
Feynman rules that can be used to obtain the diagrams, without lengthy
computations. The purpose is to check that, at least within this
subclass of diagrams, the cancellation of the prefactor of the $p^2
\omega^{D-2}$ term still occurs.

Let us describe the Feynman rules. Take for instance a term such as
$$L1\ldots  L3\ldots L2\ldots R3 \ldots  R1  \ldots R2.$$
The rules are as follows
\begin{enumerate}
\item Draw an horizontal full line and mark on it the stops (preserving
the ordering).
\item Join the corresponding L- and R-stops with a dashed line. 
\item The diagram has an incoming momentum $p$, from left to right.
\item Attach a momentum to every line (full or dashed), applying
  momentum conservation at each stop.
\item Associate a bare propagator, $G_0$, to each full line.
\item Associate a vertex function to every stop, such that its first
  argument is {\em always} the momentum runing over the dotted line.
\item For an L-stop, the second argument of the vertex will be
the momentum running over the full line to its left.
\item For an R-stop, the second argument of the vertex will be
the momentum running over the full line to its right.
\item Multiply by $1/\rho^3$ and integrate over the internal momenta.
\end{enumerate}

Applying these rules to the patterns without stop collapse, we obtain
the following contributions.\\

\paragraph{Terms $L1\ldots L2\ldots L3\ldots R3\ldots R2\ldots R1$}
\begin{fmffile}{L3-1-1}
\begin{multline}
I_1 =\parbox[b][2.2cm][b]{45mm}{
\begin{fmfgraph*}(120,50)
\fmfbottom{v1,v6}
\fmf{plain, tension=5}{v1,v2}
\fmf{plain, tension=1.2}{v2,v3}
\fmf{plain, tension=.5}{v3,v4}
\fmf{plain, tension=1.2}{v4,v5}
\fmf{plain, tension=5}{v5,v6}
\fmf{dashes, tension=1, left}{v1,v6}
\fmf{dashes, tension=1.3,left}{v2,v5}
\fmf{dashes, tension=.3, left}{v3,v4}
\end{fmfgraph*}
}=
\frac{1}{\rho^3} \int\!\! \frac{\mathrm d^D\vn{q}\mathrm d^D\vn{k} \mathrm d^D\vn{l}}
{(2\pi)^{3D} } \, V(\vn{p}-\vn{q},\vn{p}) G_0(\vn{q},z)
V(\vn{q}-\vn{k},\vn{q}) G_0(\vn{k},z)
V(\vn{k}-\vn{l},\vn{k})\times\\G_0(\vn{l},z)
 V(\vn{k}-\vn{l},\vn{k}) G_0(\vn{k},z) V(\vn{q}-\vn{k},\vn{q})
 G_0(\vn{q},z) V(\vn{p}-\vn{q},\vn{p}). 
\end{multline}
\end{fmffile}

Now we wish to compute ($S_D:$ surface of the unit-sphere in $D$
dimensions) the limit
\begin{equation}
J_1=-\frac{(2\pi)^D}{\pi S_D}\lim_{z\to 0^+} \frac{\text{Im} I_1(p,z)}{z^{(D-2)/2}},
\end{equation}
and in general, $J_k$, defined from $I_k(p,z)$ as  the same limiting
procedure.

The rules to obtain the limit painlessly are simple:
\begin{enumerate}
\item Locate a propagator, $G(q)$ whose running momentum is never
a second argument of a vertex function $V(\cdot,q)$.
\item Substitute that propagator by $-\frac{\pi S_D}{(2\pi)D}
  z^{(D-2)/2}\delta(q)$, and perform the $q$ integral.
\item Apply the simplification
\begin{equation}
G_0(\vn q,z) V(\vn q,\vn q)= \frac{V(\vn q,\vn q)}{z+V(\vn q,\vn q)} = 1 +{\cal O}(z)\,.
\end{equation}
\end{enumerate}

For $I_1$ only $l=0$ gives a contribution to $J_1$,
hence
\begin{equation}
J_1=\frac{1}{\rho^3} \int\!\! \frac{\mathrm d^D\vn{q}\, \mathrm d^D\vn{k}}
{(2\pi)^{2D} }\, V(\vn{p}-\vn{q},\vn{p}) G_0(\vn{q},z)
V(\vn{q}-\vn{k},\vn{q}) V(\vn{q}-\vn{k},\vn{q}) G_0(\vn{q},z)
V(\vn{p}-\vn{q},\vn{p}).
\end{equation}

\paragraph{Terms $L1\ldots  L2\ldots L3 \ldots R2\ldots R3\ldots R1$}
\begin{fmffile}{L3-1-2}
\begin{multline}
I_2= \parbox[b][2.2cm][b]{45mm}{
\begin{fmfgraph*}(120,50)
\fmfbottom{v1,v6}
\fmf{plain, tension=5}{v1,v2}
\fmf{plain, tension=1.2}{v2,v3}
\fmf{plain, tension=.5}{v3,v4}
\fmf{plain, tension=1.2}{v4,v5}
\fmf{plain, tension=5}{v5,v6}
\fmf{dashes, tension=1, left}{v1,v6}
\fmf{dashes, tension=.4,left}{v2,v4}
\fmf{dashes, tension=.4, left}{v3,v5}
\end{fmfgraph*}
}=
\frac{1}{\rho^3} \int\!\! \frac{\mathrm d^D\vn{q}\mathrm d^D\vn{k} \mathrm d^D\vn{l}}
{(2\pi)^{3D} }\, V(\vn{p}-\vn{q},\vn{p}) G_0(\vn{q},z) V(\vn{q}-\vn{k},\vn{q}) G_0(\vn{k},z) V(\vn{k}-\vn{l},\vn{k})\times\\G_0(\vn{l},z)
 V(\vn{q}-\vn{k},\vn{q}-\vn{k}+\vn{l}) G_0(\vn{q}-\vn{k}+\vn{l},z) V(\vn{k}-\vn{l},\vn{q}) G_0(\vn{q},z) V(\vn{p}-\vn{q},\vn{p}).
\end{multline}
\end{fmffile}
For $I_2$ one easily realizes that only $l=0$ contributes to $J_2$:
\begin{equation}
J_2=\frac{1}{\rho^3} \int\!\! \frac{\mathrm d^D\vn{q} \mathrm d^D\vn{k}}
{(2\pi)^{2D} }\, V(\vn{p}-\vn{q},\vn{p}) G_0(\vn{q},z) V(\vn{q}-\vn{k},\vn{q}) V(\vn{k},\vn{q}) G_0(\vn{q},z) V(\vn{p}-\vn{q},\vn{p}).
\end{equation}
Since $V(\vn{q}-\vn{k},\vn{q})=-V(\vn{k},\vn{q})$, one has $J_2=-J_1$.\\

\paragraph{Terms $L1\ldots  L2\ldots R2 \ldots L3\ldots R3\ldots R1$}
\begin{fmffile}{L3-1-3}
\begin{multline}
I_3=\parbox[b][2.2cm][b]{45mm}{
\begin{fmfgraph*}(120,35)
\fmfbottom{v1,v6}
\fmf{plain, tension=.8}{v1,v2}
\fmf{plain, tension=.5}{v2,v3}
\fmf{plain, tension=.8}{v3,v4}
\fmf{plain, tension=.5}{v4,v5}
\fmf{plain, tension=.8}{v5,v6}
\fmf{dashes, tension=2, left}{v1,v6}
\fmf{dashes, tension=.2,left}{v2,v3}
\fmf{dashes, tension=.2, left}{v4,v5}
\end{fmfgraph*}
}=
\frac{1}{\rho^3} \int\!\! \frac{\mathrm d^D\vn{q}\mathrm d^D\vn{k} \mathrm d^D\vn{l}}
{(2\pi)^{3D} }\, V(\vn{p}-\vn{q},\vn{p}) G_0(\vn{q},z) V(\vn{q}-\vn{k},\vn{q}) G_0(\vn{k},z) V(\vn{q}-\vn{k},\vn{q})\times\\G_0(\vn{q},z)
 V(\vn{q}-\vn{l},\vn{q}) G_0(\vn{l},z) V(\vn{q}-\vn{l},\vn{q}) G_0(\vn{q},z) V(\vn{p}-\vn{q},\vn{p}).
\end{multline}
\end{fmffile}
Both $\vn{k}=0$ and $\vn{l}=0$ contribute to $J_3$ (for $\vn{k}=0$ we
changed the dummy variable $\vn{l}$ to $\vn{k}$):
\begin{equation}
J_3=\frac{2}{\rho^3} \int\!\! \frac{\mathrm d^D\vn{q}\,\mathrm d^D\vn{k}}
{(2\pi)^{2D} }\, V(\vn{p}-\vn{q},\vn{p}) V(\vn{q}-\vn{k},\vn{q}) G_0(\vn{k},z) V(\vn{q}-\vn{k},\vn{q}) G_0(\vn{q},z) V(\vn{p}-\vn{q},\vn{p}).
\end{equation}

\paragraph{Terms $L1\ldots L3\ldots R3 \ldots L2\ldots R1\ldots R2$}
\begin{fmffile}{L3-2-1}
\begin{multline}
I_4=\parbox[b][1.5cm][b]{45mm}{
\begin{fmfgraph*}(120,35)
\fmfstraight
\fmfbottom{v1,v2,v3,v4,v5,v6}
\fmf{plain, tension=1}{v1,v2}
\fmf{plain, tension=1}{v2,v3}
\fmf{plain, tension=1}{v3,v4}
\fmf{plain, tension=1}{v4,v5}
\fmf{plain, tension=1}{v5,v6}
\fmf{dashes, tension=2, left}{v1,v5}
\fmf{dashes, tension=.2,left}{v2,v3}
\fmf{dashes, tension=.5, left}{v4,v6}
\end{fmfgraph*}
}=
\frac{1}{\rho^3} \int\!\! \frac{\mathrm d^D\vn{q}\mathrm d^D\vn{k} \mathrm d^D\vn{l}}
{(2\pi)^{3D} }\, V(\vn{p}-\vn{q},\vn{p}) G_0(\vn{q},z) V(\vn{q}-\vn{k},\vn{q}) G_0(\vn{k},z) V(\vn{q}-\vn{k},\vn{q})\times\\G_0(\vn{q},z)
 V(\vn{q}-\vn{l},\vn{q}) G_0(\vn{l},z) V(\vn{p}-\vn{q},\vn{p}-\vn{q}+\vn{l}) G_0(\vn{p}-\vn{q}+\vn{l},z) V(\vn{q}-\vn{l},\vn{p}).
\end{multline}
\end{fmffile}

For $J_4$ both $\vn{k}=0$ and $\vn{l}=0$ are relevant:
\begin{multline}
J_4=\frac{1}{\rho^3} \int\!\! \frac{\mathrm d^D\vn{q}\,\mathrm d^D\vn{k}}
{(2\pi)^{2D} }\, V(\vn{p}-\vn{q},\vn{p}) V(\vn{q}-\vn{k},\vn{q}) G_0(\vn{k},z) V(\vn{p}-\vn{q},\vn{p}-\vn{q}+\vn{k}) G_0(\vn{p}-\vn{q}+\vn{k},z) V(\vn{q}-\vn{k},\vn{p}) +\\
\frac{1}{\rho^3} \int\!\! \frac{\mathrm d^D\vn{q}\,\mathrm d^D\vn{k}}
{(2\pi)^{2D} }\, V(\vn{p}-\vn{q},\vn{p}) G_0(\vn{q},z) V(\vn{q}-\vn{k},\vn{q}) G_0(\vn{k},z) V(\vn{q}-\vn{k},\vn{q}) V(\vn q,\vn{p}).
\end{multline}

\paragraph{Terms $L1\ldots  L2\ldots L3\ldots R3 \ldots  R1\ldots R2$}
\begin{fmffile}{L3-2-2}
\begin{multline}
I_5=\parbox[b][1.5cm][b]{45mm}{
\begin{fmfgraph*}(120,35)
\fmfstraight
\fmfbottom{v1,v2,v3,v4,v5,v6}
\fmf{plain, tension=1}{v1,v2}
\fmf{plain, tension=1}{v2,v3}
\fmf{plain, tension=.5}{v3,v4}
\fmf{plain, tension=1}{v4,v5}
\fmf{plain, tension=1}{v5,v6}
\fmf{dashes, tension=2, left}{v1,v5}
\fmf{dashes, tension=.2,left}{v2,v6}
\fmf{dashes, tension=.5, left}{v3,v4}
\end{fmfgraph*}
}
=\frac{1}{\rho^3} \int\!\! \frac{\mathrm d^D\vn{q}\mathrm d^D\vn{k} \mathrm d^D\vn{l}}
{(2\pi)^{3D} }\, V(\vn{p}-\vn{q},\vn{p}) G_0(\vn{q},z) V(\vn{q}-\vn{k},\vn{q}) G_0(\vn{k},z) V(\vn{k}-\vn{l},\vn{k})\times\\
G_0(\vn{l},z) V(\vn{k}-\vn{l},\vn{k}) G_0(\vn{k},z) V(\vn{p}-\vn{q},\vn{p}-\vn{q}+\vn{k}) G_0(\vn{p}-\vn{q}+\vn{k},z) V(\vn{q}-\vn{k},\vn{p}).
\end{multline}
\end{fmffile}

The only relevant contributions is now $\vn{l}=0$:
\begin{equation}
J_5=\frac{1}{\rho^3} \int\!\! \frac{\mathrm d^D\vn{q}\,\mathrm d^D\vn{k}}
{(2\pi)^{2D} }\, V(\vn{p}-\vn{q},\vn{p}) G_0(\vn{q},z) V(\vn{q}-\vn{k},\vn{q})V(\vn{p}-\vn{q},\vn{p}-\vn{q}+\vn{k}) G_0(\vn{p}-\vn{q}+\vn{k},z) V(\vn{q}-\vn{k},\vn{p}).
\end{equation}

\paragraph{Terms $L1\ldots  L2\ldots R1 \ldots L3\ldots R3 \ldots R2$}
\begin{fmffile}{L3-2-3}
\begin{multline}
I_6=\parbox[b][1.5cm][b]{45mm}{
\begin{fmfgraph*}(120,35)
\fmfstraight
\fmfbottom{v1,v2,v3,v4,v5,v6}
\fmf{plain, tension=7}{i1,v1}
\fmf{plain, tension=1}{v1,v2}
\fmf{plain, tension=1}{v2,v3}
\fmf{plain, tension=.5}{v3,v4}
\fmf{plain, tension=1}{v4,v5}
\fmf{plain, tension=1}{v5,v6}
\fmf{plain, tension=7}{v6,o1}
\fmf{dashes, tension=2, left}{v1,v3}
\fmf{dashes, tension=.2,left}{v2,v6}
\fmf{dashes, tension=.5, left}{v4,v5}
\end{fmfgraph*}
}=
\frac{1}{\rho^3} \int\!\! \frac{\mathrm d^D\vn{q}\mathrm d^D\vn{k} \mathrm d^D\vn{l}}
{(2\pi)^{3D} }\, V(\vn{p}-\vn{q},\vn{p}) G_0(\vn{q},z) V(\vn{q}-\vn{k},\vn{q}) G_0(\vn{k},z) V(\vn{p}-\vn{q},\vn{p}-\vn{q}+\vn{k})\times\\
G_0(\vn{p}-\vn{q}+\vn{k},z) V(\vn{p}-\vn{q}+\vn{k}-\vn{l},\vn{p}-\vn{q}+\vn{k}) G_0(\vn{l},z) V(\vn{p}-\vn{q}+\vn{k}-\vn{l},\vn{p}-\vn{q}+\vn{k}) G_0(\vn{p}-\vn{q}+\vn{k},z) V(\vn{q}-\vn{k},\vn{p}).
\end{multline}
\end{fmffile}

$J_6$ stems both from $\vn{k}=0$ and from $\vn{l}=0$. For the
$\vn{k}=0$ contribution, 
we make the change of variable $\vn{q}\longrightarrow \vn{p}-\vn{q}$ 
to identify the cancellation with $J_3$:
\begin{multline}
J_6=\frac{1}{\rho^3} \int\!\! \frac{\mathrm d^D\vn{q} \,\mathrm d^D\vn{k}}
{(2\pi)^{2D} }\, V(\vn q,\vn{p}) V(\vn{q}-\vn{k},\vn{q}) G_0(\vn{k},z) V(\vn{q}-\vn{k},\vn{q}) G_0(\vn{q},z) V(\vn{p}-\vn{q},\vn{p})+\\
\frac{1}{\rho^3} \int \frac{\mathrm{d}^Dq \mathrm{d}^D\vn{k}}
{(2\pi)^{2D} }\ V(\vn{p}-\vn{q},\vn{p}) G_0(\vn{q},z) V(\vn{q}-\vn{k},\vn{q}) G_0(\vn{k},z) V(\vn{p}-\vn{q},\vn{p}-\vn{q}+\vn{k}) V(\vn{q}-\vn{k},\vn{p}).
\end{multline}

\paragraph{Terms $L1\ldots  L2\ldots L3\ldots R1 \ldots R3 \ldots R2$}
\begin{fmffile}{L3-3-1}
\begin{multline}
I_7=\parbox[b][1.3cm][b]{38mm}{
\begin{fmfgraph*}(100,35)
\fmfstraight
\fmfbottom{v1,v2,v3,v4,v5,v6}
\fmf{plain, tension=1}{v1,v2}
\fmf{plain, tension=1}{v2,v3}
\fmf{plain, tension=.5}{v3,v4}
\fmf{plain, tension=1}{v4,v5}
\fmf{plain, tension=1}{v5,v6}
\fmf{dashes, tension=2, left}{v1,v4}
\fmf{dashes, tension=.2,left}{v2,v6}
\fmf{dashes, tension=.5, left}{v3,v5}
\end{fmfgraph*}
}=
\frac{1}{\rho^3} \int\!\! \frac{\mathrm d^D\vn{q}\mathrm d^D\vn{k} \mathrm d^D\vn{l}}
{(2\pi)^{3D} }\, V(\vn{p}-\vn{q},\vn{p}) G_0(\vn{q},z) V(\vn{q}-\vn{k},\vn{q}) G_0(\vn{k},z) V(\vn{k}-\vn{l},\vn{k})\times\\
G_0(\vn{l},z) V(\vn{p}-\vn{q},\vn{p}-\vn{q}+\vn{l}) G_0(\vn{p}-\vn{q}+\vn{l},z) V(\vn{k}-\vn{l},\vn{p}-\vn{q}+\vn{k}) G_0(\vn{p}-\vn{q}+\vn{k},z) V(\vn{q}-\vn{k},\vn{p}).
\end{multline}
\end{fmffile}

Only $\vn{l}=0$ contributes to $J_7$:
\begin{equation}
J_7=\frac{1}{\rho^3} \int\!\! \frac{\mathrm d^D\vn{q}\,\mathrm d^D\vn{k}}
{(2\pi)^{2D} }\, V(\vn{p}-\vn{q},\vn{p})
G_0(\vn{q},z)V(\vn{q}-\vn{k},\vn{q}) V(\vn{k},\vn{p}-\vn{q}+\vn{k})
G_0(\vn{p}-\vn{q}+\vn{k},z) V(\vn{q}-\vn{k},\vn{p}).
\end{equation}

\paragraph{Terms $L1\ldots  L3\ldots L2\ldots  R1 \ldots R3 \ldots R2$}
\begin{fmffile}{L3-3-2}
\begin{multline}
I_8=\parbox[b][1cm][b]{38mm}{
\begin{fmfgraph*}(100,35)
\fmfstraight
\fmfbottom{v1,v2,v3,v4,v5,v6}
\fmf{plain, tension=1}{v1,v2}
\fmf{plain, tension=1}{v2,v3}
\fmf{plain, tension=.5}{v3,v4}
\fmf{plain, tension=1}{v4,v5}
\fmf{plain, tension=1}{v5,v6}
\fmf{dashes, tension=2, left}{v1,v4}
\fmf{dashes, tension=.2,left}{v2,v5}
\fmf{dashes, tension=.5, left}{v3,v6}
\end{fmfgraph*}
}=
\frac{1}{\rho^3} \int\!\! \frac{\mathrm d^D\vn{q}\mathrm d^D\vn{k} \mathrm d^D\vn{l}}
{(2\pi)^{3D} }\, V(\vn{p}-\vn{q},\vn{p}) G_0(\vn{q},z) V(\vn{q}-\vn{k},\vn{q}) G_0(\vn{k},z) V(\vn{k}-\vn{l},\vn{k})\times\\G_0(\vn{l},z)V(\vn{p}-\vn{q},\vn{p}-\vn{q}+\vn{l}) G_0(\vn{p}-\vn{q}+\vn{l},z) V(\vn{q}-\vn{k},\vn{p}+\vn{l}-\vn{k}) G_0(\vn{p}+\vn{l}-\vn{k},z) V(\vn{k}-\vn{l},\vn{p}).
\end{multline}
\end{fmffile}

Again, only $\vn{l}=0$ matters:
\begin{equation}
J_8=\frac{1}{\rho^3} \int\!\! \frac{\mathrm d^D\vn{q}\,\mathrm d^D\vn{k}}
{(2\pi)^{2D} }\, V(\vn{p}-\vn{q},\vn{p}) G_0(\vn{q},z) V(\vn{q}-\vn{k},\vn{q}) V(\vn{q}-\vn{k},\vn p-\vn{k}) G_0(\vn{p}-\vn{k},z) V(\vn{k},\vn{p})\,.
\end{equation}

\paragraph{Terms $L1\ldots  L3\ldots L2\ldots R3 \ldots  R1  \ldots R2$}
\begin{fmffile}{L3-3-3}
\begin{multline}
I_9=\parbox[b][1.3cm][b]{38mm}{
\begin{fmfgraph*}(100,35)
\fmfstraight
\fmfbottom{v1,v2,v3,v4,v5,v6}
\fmf{plain, tension=1}{v1,v2}
\fmf{plain, tension=1}{v2,v3}
\fmf{plain, tension=.5}{v3,v4}
\fmf{plain, tension=1}{v4,v5}
\fmf{plain, tension=1}{v5,v6}
\fmf{dashes, tension=2, left}{v1,v5}
\fmf{dashes, tension=.2,left}{v2,v4}
\fmf{dashes, tension=.5, left}{v3,v6}
\end{fmfgraph*}
}=
\frac{1}{\rho^3} \int\!\! \frac{\mathrm d^D\vn{q}\mathrm d^D\vn{k} \mathrm d^D\vn{l}}
{(2\pi)^{3D} }\, V(\vn{p}-\vn{q},\vn{p}) G_0(\vn{q},z) V(\vn{q}-\vn{k},\vn{q}) G_0(\vn{k},z) V(\vn{k}-\vn{l},\vn{k})\times\\G_0(\vn{l},z) V(\vn{q}-\vn{k},\vn{q}-\vn{k}+\vn{l}) G_0(\vn{q}-\vn{k}+\vn{l},z) V(\vn{p}-\vn{q},\vn{p}-\vn{k}+\vn{l}) G_0(\vn{p}-\vn{k}+\vn{l},z) V(\vn{k}-\vn{l},\vn{p}).
\end{multline}
\end{fmffile}

And, once again, only $\vn{l}=0$ contributes:
\begin{equation}
J_9=\frac{1}{\rho^3} \int\!\! \frac{\mathrm d^D\vn{q}\,\mathrm d^D\vn{k}}
{(2\pi)^{2D} }\, V(\vn{p}-\vn{q},\vn{p}) G_0(\vn{q},z) V(\vn{q}-\vn{k},\vn{q}) V(\vn{p}-\vn{q},\vn p-\vn{k}) G_0(\vn{p}-\vn{k},z) V(\vn{k},\vn{p}).
\end{equation}

\paragraph{Terms $L1\ldots   L3\ldots R1 \ldots L2\ldots R3 \ldots R2$}
\begin{fmffile}{L3-4-1}
\begin{multline}
I_{10}=\parbox[b][1cm][b]{38mm}{
\begin{fmfgraph*}(100,35)
\fmfstraight
\fmfbottom{v1,v2,v3,v4,v5,v6}
\fmf{plain, tension=1}{v1,v2}
\fmf{plain, tension=1}{v2,v3}
\fmf{plain, tension=.5}{v3,v4}
\fmf{plain, tension=1}{v4,v5}
\fmf{plain, tension=1}{v5,v6}
\fmf{dashes, tension=2, left}{v1,v3}
\fmf{dashes, tension=.2,left}{v2,v5}
\fmf{dashes, tension=.5, left}{v4,v6}
\end{fmfgraph*}
}=
\frac{1}{\rho^3} \int\!\! \frac{\mathrm d^D\vn{q}\mathrm d^D\vn{k} \mathrm d^D\vn{l}}
{(2\pi)^{3D} }\, V(\vn{p}-\vn{q},\vn{p}) G_0(\vn{q},z) V(\vn{q}-\vn{k},\vn{q}) G_0(\vn{k},z) V(\vn{p}-\vn{q},\vn{p}-\vn{q}+\vn{k})\times\\G_0(\vn{p}-\vn{q}+\vn{k},z) V(\vn{p}-\vn{q}+\vn{k}-\vn{l},\vn{p}-\vn{q}+\vn{k}) G_0(\vn{l},z) V(\vn{q}-\vn{k},\vn{q}-\vn{k}+\vn{l}) G_0(\vn{q}-\vn{k}+\vn{l},z) V(\vn{p}-\vn{q}+\vn{k}-\vn{l},\vn{p}).
\end{multline}
\end{fmffile}

Here we have a contribution from $\vn{k}=0$ as well as from $\vn{l}=0$:
\begin{equation}
J_{10}=\frac{2}{\rho^3} \int \!\!\frac{\mathrm d^D\vn{q}\,\mathrm d^D\vn{k}}
{(2\pi)^{2D} }\, V(\vn{p}-\vn{q},\vn{p}) V(\vn{p}-\vn{q}-\vn{k},\vn{p}-\vn{q}) G_0(\vn{k},z) V(\vn q,\vn{q}+\vn{k}) G_0(\vn{q}+\vn{k},z) V(\vn{p}-\vn{q}-\vn{k},\vn{p}).
\end{equation}

\subsubsection{Resummation of the imaginary parts}
The resummation of the imaginary parts of the previous diagrams is
simple. Using the properties of the vertex $V(\vn{p},\vn{q})$ and
changing carefully the integration variables when necessary we can
show that
\begin{align}
J_1+J_2&=0\\
J_3+J_4+J_6+J_{10}&=0\\
J_5+J_7&=0\\
J_8+J_9&=0
\end{align}
so that
\begin{equation}
\sum_{i=1}^{10}J_i=0,
\end{equation}
and the total contribution to the imaginary part proportional to
$z^{(D-2)/2}$ vanishes.


\section{A field theory approach}\label{FIELD}

In this section we will introduce a field-theoretical representation
for the resolvent $G(\vn p,z)$.  Within this formalism one is able to
obtain the perturbative computation for the self-energy in a more
strightforward way than with previous
formulations~\cite{ERM1}. Interestingly enough, due to the ultraviolet
behaviour of the bare propagator of the field involved, such
perturbative expansion yields some divergent terms that can be summed
up to zero. The starting point is the following representation for the
resolvent:
\begin{equation}
G(\vn{p},z)=\overline{\frac{1}{N}\sum_{ij}e^{i\vn p\cdot (\vn x_i-\vn
    x_j)}\frac{\int \left( \prod_i^N d\phi_i\right) \phi_i \phi_j
    \exp\left\{-\frac{1}{2}\sum_{lm}\phi_l \left[\left(z-\sum_k f(\vn
      x_l-\vn x_k)\right)\delta_{lm} +f(\vn x_l-\vn x_m)\right]\phi_m
    \right\}}{\int \left( \prod_i^N d\phi_i\right)
    \ \exp\left\{-\frac{1}{2}\sum_{lm}\phi_l \left[\left(z-\sum_k
      f(\vn x_l-\vn x_k)\right)\delta_{lm} +f(\vn x_l-\vn
      x_m)\right]\phi_m \right\}}}.
\end{equation} 
Introducing the fields
\begin{align}
\phi(\vn{x}) &\equiv
\begin{cases}
\phi_i & \vn{x}=\vn{x}_i, \\
\text{arbitrary} & \vn{x} \neq \vn{x}_i,
\end{cases} \\
\rho(\vn x)&\equiv\frac{1}{\rho}\sum_k\delta(\vn x-\vn x_k),
\end{align} 
one has
\begin{equation}
G(\vn p,z)=\overline{\frac{\rho^2}{N}\int\mathrm{d}^D \vn{x}
  \mathrm{d}^D \vn{y}\ e^{\mathrm{i} \vn{p} \cdot (\vn{x}-\vn{y})}
  \frac{\rho(\vn{x})\rho(\vn{y})}{\mathcal{Z}_\rho}\int \left(\prod_i^N
  d\phi(\vn x_i) \phi(\vn{x})\phi(\vn{y})\ \exp\left\{ S_\rho\left[ \phi \right]
  \right\}\right)},
\end{equation} 
where we have introduced the action and the partition function at a
fixed realization of the disorder, given respectively by
\begin{align}
S_\rho\left[ \phi \right]&=-\frac{\rho}{2}\int\!\!
 \mathrm{d}^D\vn x\,\mathrm{d}^D\vn y\, \phi(\vn{x})\left[
  z\rho(\vn{x})\delta(\vn{x}-\vn{y})-
  \rho\rho(\vn{x})\delta(\vn{x}-\vn{y}) \int\!\! \mathrm{d}^D\vm{\sigma}\,
  f(\vn{x}-\vm{\sigma})\rho(\vm{\sigma}) +
  \rho\rho(\vn{x})f(\vn{x}-\vn{y})\rho(\vn{y})\right]\phi(\vn{y}), \label{eq:action}\\
Z_\rho&=\int\!\! \left( \prod_i^N \mathrm{d}\phi(\vn x_i) \right)\, \exp\left\{
S_\rho\left[ \phi \right] \right\}.
\end{align} 

Now we note that the action Eq.~(\ref{eq:action}) depends on the field
$\phi$ only through the values that it assumes on the random positions
$\left\{\vn x_i\right\}$. In fact, in the action, the field $\phi$ is
always multiplied by the random field $\rho$, which selects the random
points of the lattice $\left\{\vn x_i\right\}$. So, we can substitute
the discretized functional measure with the continuous one: this is a
crucial step. The continuous version of the fuctional integral is
invariant under the following transformation of the field $\phi$,
which we shall call gauge transformation:
\begin{equation}
\phi'(\vn{x})=\phi(\vn{x})+h(\vn{x}), \qquad \text{with}\quad
 h(\vn x_i)=0, \quad i=1\ldots N. \label{eq:condition}
\end{equation} 
This is a local transformation, but we can see that its global version
is trivial because the condition Eq.~(\ref{eq:condition}) implies that a
global transformation can be possible only for $h=0$. This local symmetry is not
present in other field-theoretic formulations~\cite{ERM1}.

We now look at the resolvent: it can be written in the form
\begin{equation}
G(\vn p,z)=\int\!\!\mathrm{d}^D\vn{x}\mathrm{d}^D\vn{y}\, e^{i\vn p \cdot
  (\vn{x}-\vn{y})}\ \overline{\frac{\rho^2}{N}
  \rho(\vn{x})\rho(\vn{y})\left\langle\phi(\vn{x})\phi(\vn{y})\right\rangle},
\end{equation}
where $\left\langle\cdot\right\rangle$ stands for the average over the action
$S_\rho[\phi]$. We immediately see that
$\rho(\vn{x})\rho(\vn{y})\big\langle \phi(\vn{x})\phi(\vn{y})\big\rangle$ is gauge
invariant. With the change of variables
\begin{equation}
\rho(\vn{x})=1+\delta\rho(\vn{x})
\end{equation}
the resolvent can be written
\begin{equation}
G(\vn p,z)=\int\!\!\mathrm{d}^D\vn{x}\mathrm{d}^D\vn{y}\, e^{i\vn
  p\cdot(\vn{x}-\vn{y})}\left\{ \overline{\frac{\rho^2}{N}
  \left<\phi(\vn{x})\phi(\vn{y})\right>}+2\overline{\frac{\rho^2}{N}
  \delta\rho(\vn{x})\left<\phi(\vn{x})\phi(\vn{y})\right>}+\overline{\frac{\rho^2}{N}
  \delta\rho(\vn{x})
  \delta\rho(\vn{y})\left<\phi(\vn{x})\phi(\vn{y})\right>}
\right\},
\label{eq:G_deltarho}
\end{equation} 
with the action
\begin{equation}
\begin{split}
S_\rho [\phi]&=-\frac{\rho}{2}\int\!\!\mathrm{d}^D\vn{x}\mathrm{d}^D\vn{y}\, \phi(\vn{x}) \left[
  (z-\rho\tilde f (0))\delta(\vn{x}-\vn{y})+\rho
  f(\vn{x}-\vn{y})\right]\phi(\vn{y})+  \\
&\quad -\frac{\rho}{2}\int\!\! \mathrm{d}^D\vn{x}\mathrm{d}^D\vn{y}d\vm{\sigma}\,
 \phi(\vn{x})\phi(\vn{y})\delta\rho(\vm{\sigma})
  V_3(\vn{x},\vn{y},\vm{\sigma})  \\
&\quad -\frac{\rho}{2}\int\!\!\mathrm{d}^D\vn{x}\mathrm{d}^D\vn{y}d\vm{\sigma}d\vm{\gamma}\,
 \phi(\vn{x})\phi(\vn{y})\delta\rho(\vm{\sigma})\delta\rho(\vm{\gamma})
V_4(\vn{x},\vn{y},\vm{\sigma},\vm{\gamma}),
\end{split}
\label{ACTION}
\end{equation}
where 
\begin{align}
V_3(\vn{x},\vn{y},\vm{\sigma})&=\left[(z-\rho\hat f
  (\vn 0))\delta(\vn{x}-\vn{y})\delta(\vm{\sigma} -
  \vn{x})-\rho\delta(\vn{x}-\vn{y})f(\vn{x}-\vm{\sigma})-\rho
  f(\vn{x}-\vn{y})\left(\delta(\vn{x}-\vm{\sigma})+\delta(\vm{\sigma}-\vn{y})
  \right)\right],\\
V_4(\vn{x},\vn{y},\vm{\sigma},\vm{\gamma})&=\rho\delta(\vm{\gamma}-\vn{x})f(\vn{x}-\vn{y})\delta(\vn{y}-\vm{\sigma})-\rho\delta(\vm{\gamma}-\vn{x})\delta(\vn{x}-\vn{y})f(\vn{x}-\vm{\sigma}).
\end{align}
Note that the first term of Eq.~(\ref{eq:G_deltarho}), when computed in
the limit $\delta \rho =0$ of the action (Eq.~(\ref{ACTION})) yields the
bare propagator $G_0(\vn p,z)$. The fist term then corresponds to the
free (gaussian) part of the field theory and the terms involving three
and four fields are the interacting part. The latter can be treated
perturbatively using standard diagrammatic techniques of field
theory. One can easily see from the form of the interacting terms that
in such diagrams no loops involving the $\delta \rho$ field may arise
because, at fixed disorder, it acts as an external field while a generic $n$-loop
diagram comes from the average over the disorder and yields a $1/\rho^n$ contribution to the resolvent.

\subsection{The correlation functions for the density field}

In order to perform the loop expansion one needs the $n$-point
correlation functions of the $\delta \rho$ field. Since
\begin{equation}
\overline{\rho(\vn{x})}=1+\overline{\delta\rho(\vn{x})}=\int\!\!
 \left( \prod_i^N\frac{\mathrm{d}^D\vn{x}_i}{V} \right)\frac{1}{\rho}\sum_k^N \delta(\vn{x}-\vn{x}_k)=\frac{1}{\rho}\frac{N}{V}=1,
\end{equation}
one has
\begin{equation}
\overline{\delta\rho(\vn{x})}=0.\label{eq:delta-rho-1-points}
\end{equation} 
Similarly, the fact that
\begin{equation}
\begin{split}
\overline{\rho(\vn{x})\rho(\vn{y})}&=1+\overline{\delta\rho(\vn{x})}+
\overline{\delta\rho(\vn{y})}+\overline{\delta\rho(\vn{x})\delta\rho(\vn{y})}=1+\overline{\delta\rho(\vn{x})\delta\rho(\vn{y})}=\\
&=\frac{1}{\rho^2}\sum_k^N\sum_j^N \int\!\! \left(\prod_i^N \frac{\mathrm{d}^D\vn{x}_i}{V}\right)\, \delta(\vn{x}-\vn{x}_k)\delta(\vn{y}-\vn{x}_j)=\\
&=\frac{1}{\rho^2}\sum_{k\neq j} \int\!\! \left(\prod_i^N \frac{\mathrm{d}^D\vn{x}_i}{V}\right)\, \delta(\vn{x}-\vn{x}_k)\delta(\vn{y}-\vn{x}_j)+\frac{1}{\rho^2}\sum_k^N \int \left(\prod_i^N \frac{\mathrm{d}^D\vn{x}_i}{V}\right) \delta(\vn{x}-\vn{x}_k)\delta(\vn{y}-\vn{x}_k)\\
&=\frac{N(N-1)}{\rho^2 V^2}+\frac{N}{V
  \rho^2}\delta(\vn{x}-\vn{y})\longrightarrow 1+\frac{1}{\rho} \delta(\vn{x}-\vn{y})
\end{split}
\end{equation}
implies that
\begin{equation}
\overline{\delta\rho(\vn{x})\delta\rho(\vn{y})}=\frac{1}{\rho}
\delta(\vn{x}-\vn{y}).
\label{eq:delta-rho-2-points}
\end{equation} 
To carry our the perturbative expansion up to order $1/\rho^2$, the 3-
and 4-point correlations are needed. These can be derived according to the lines
described above, giving
\begin{align}
\label{eq:delta-rho-3-points}
\overline{\delta\rho(\vn{x})\delta\rho(\vn{y})\delta\rho(\vn{z})}&=\frac{1}{\rho^2}\delta(\vn{x}-\vn{y})\delta(\vn{y}-\vn{z})=\overline{\delta\rho(\vn{x})\delta\rho(\vn{y})}\cdot
\overline{\delta\rho(\vn{y})\delta\rho(\vn{z})},\\ 
\label{eq:delta-rho-4-points}
\overline{\delta\rho(\vn{x})\delta\rho(\vn{y})\delta\rho(\vn{z})\delta\rho(\vn{t})}&=\frac{1}{\rho^3}\delta(\vn{x}-\vn{y})\delta(\vn{y}-\vn{z})\delta(\vn{z}-\vn{t})+\overline{\delta\rho(\vn{x})\delta\rho(\vn{y})}\cdot
\overline{\delta\rho(\vn{z})\delta\rho(\vn{t})}+\\\nonumber
&\quad\overline{\delta\rho(\vn{x})\delta\rho(\vn{z})}\cdot
\overline{\delta\rho(\vn{y})\delta\rho(\vn{t})}+
\overline{\delta\rho(\vn{x})\delta\rho(\vn{t})}\cdot
\overline{\delta\rho(\vn{y})\delta\rho(\vn{z})}.
\end{align}

\subsubsection{The general expression}
We may write as well the expression for the arbitrary $n$-point
correlation $\overline{\delta\rho(\vn{y}_1)
  \delta\rho(\vn{y}_2)\ldots\delta\rho(\vn{y}_k)}$, needed to compute
the self-energy to order $1/\rho^3$ or higher in the field theory.  To
give our result, we shall need some notations.

Let $\omega$ be an
arbitrary partition of the set $\{1,2,\ldots,k\}$ into subsets.  For
instance, for $k=4$, $\omega$ could be a partition into two subsets,
such as $\omega=\{\,\{1,2\},\{3,4\}\}$ or
$\omega=\{\,\{1,3\},\{2,4\}\}$, or a partition into 4 subsets such as
$\{\,\{1\},\{2\},\{3\},\{4\}\,\}$, etc. Let
$\left\Vert\alpha\right\Vert$ be the cardinality of the set $\alpha$,
for instance, if $\omega=\{\,\{1,2\},\{3,4\}\}$, then
$\Vert\omega\Vert=2\,.$ 

We also define ${\cal P}^{(k)}$, the set of all possible partitions of
$\{1,2,\ldots,k\}$. Given a partition $\omega$, the subsets associated
to it will be $\Omega_{l,\omega}$, with
$l=1,2,\ldots,\Vert\omega\Vert\,.$ We shall need to consider ${\cal
  H}^{(k)}$, a subset of the set of all partitions ${\cal P}^{(k)}$.
${\cal H}^{(k)}$ is made of all partitions $\omega$ such that
$\left\Vert\Omega_{l,\omega}\right\Vert>1$ for all
$l=1,2,\ldots,\left\Vert \omega\right\Vert$, i.e.\ partitions in which
none of the subsets contains less than {\em two} integers.  Then the
general result is:
\begin{equation}\label{eq:delta-rho-k-points}
\overline{\delta\rho(\vn{y}_1)
  \delta\rho(\vn{y}_2)\ldots\delta\rho(\vn{y}_k)}=\sum_{\omega\in {\cal H}^{(k)}}
\frac{1}{\rho^{k-\left\Vert\omega\right\Vert}}\left[\prod_{l=1}^{\left\Vert\omega\right\Vert}\left(\prod_{r=1}^{\left\Vert\Omega_{l,\omega}\right\Vert-1} \delta\bigl(\vn{y}_{\alpha_r^{(l,\omega)}}-\vn{y}_{\alpha_{r+1}^{(l,\omega)}}\bigr)\right)\right].
\end{equation}
The proof is given in app.~\ref{AP:PROOF}. To recover
Eq.~(\ref{eq:delta-rho-3-points}) from this formula, note
that the set ${\cal H}^{(3)}$ of allowed partitions of
$\{\vn{x},\vn{y},\vn{z}\}$ contains a single partition, with just
one subset ($\left\Vert\omega\right\Vert =1$), namely
$\omega=\bigl\{\{\vn{x},\vn{y},\vn{z}\}\bigr\}$. On the other hand, to obtain
Eq.~(\ref{eq:delta-rho-4-points})  we need the set ${\cal H}^{(4)}$ of allowed
partitions for $\bigl\{\{\vn{x},\vn{y},\vn{z},\vn{t}\}\bigr\}$. There are four
such partitions, namely $\omega_1=\{\vn{x},\vn{y},\vn{z},\vn{t}\}$,
$\omega_2=\bigl\{\,\{\vn{x},\vn{y}\},\{\vn{z},\vn{t}\}\,\bigr\}$, 
$\omega_3=\bigl\{\,\{\vn{x},\vn{z}\},\{\vn{y},\vn{t}\}\,\bigr\}$, and 
$\omega_4=\bigl\{\,\{\vn{x},\vn{t}\},\{\vn{y},\vn{z}\}\,\bigr\}$. Clearly,
$\left\Vert\omega_1\right\Vert=1$, while $\left\Vert\omega_2\right\Vert=\left\Vert\omega_3\right\Vert=\left\Vert\omega_4\right\Vert=2$.

\subsection{Diagrammatic expansion: one loop}

In order to write down the one-loop term it turns out to be convenient
to write $V_3$ and $V_4$ in terms of the interaction vertex
(Eq.~(\ref{VERTEX})):
\begin{align}
\int\!\!\mathrm{d}^D\vn{x}\mathrm{d}^D\vm{\sigma}\,e^{i \vn{p}_1 \cdot \vn{x}+i\vn{p}_2\cdot
  \vm{\sigma}} V_3(\vn{x},\vn{y},\vm{\sigma})&=\left[G_0^{-1}(\vn{p}_1)
  - V(\vn{p}_2,-\vn{p}_1)\right]e^{i(\vn{p}_1+\vn{p}_2)\cdot\vn{y}} \equiv \mu
  (\vn{p}_1,\vn{p}_2)e^{i(\vn{p}_1+\vn{p}_2)\cdot \vn{y}}\\ 
\int\!\!\mathrm{d}^D\vn{x}\mathrm{d}^D\vm{\sigma}\mathrm{d}^D\vm{\gamma}\,
e^{i\vn{p}_1\cdot\vn{x}+i\vn{p}_2\cdot \vm{\sigma}+i\vn{p}_3\cdot\vm{\gamma}}
V_4(\vn{x},\vn{y},\vm{\sigma},\vm{\gamma})&=-V(\vn{p}_2,-\vn{p}_1)e^{i(\vn{p}_1+\vn{p}_2+
  \vn{p}_3)\cdot\vn{y}}.
\end{align}
The latter expression depends only on two momenta. Thus when the vertex
$V_4$ is involved, one has to make its expression symmetric by joining
the $\delta\rho$ propagators with the two possible external links
offered by this vertex. 

\begin{figure}
\begin{fmffile}{notazione}
\begin{align*}
\parbox{35mm}{
\begin{fmfgraph*}(80,30)
\fmfleft{i1}
\fmfright{o1}
\fmf{plain, label=$\vn p$}{i1,o1}
\end{fmfgraph*}
}&= \frac{1}{\rho}G_0(\vn p,z)\\
\parbox{35mm}{
\begin{fmfgraph*}(80,40)
\fmfleft{i1}
\fmfright{o1}
\fmf{photon, label=$\vn p$}{i1,o1}
\end{fmfgraph*}
}&= \frac{1}{\rho}\\
\parbox{35mm}{
\begin{fmfgraph*}(80,60)
\fmfleft{i1,i2}
\fmfright{o1}
\fmf{plain_arrow, label=$\vn p+\vn q$}{v1,o1}
\fmf{plain_arrow, label=$\vn p$}{i1,v1}
\fmf{photon, label=$\vn q$}{i2,v1}
\fmfv{d.shape=circle, d.size=3thick}{v1}
\end{fmfgraph*}
}&=\mu(\vn p,\vn q)\\
\parbox{35mm}{
\begin{fmfgraph*}(80,70)
\fmfleft{i1,i2,i3}
\fmfright{o1}
\fmf{plain_arrow, tension=3, label=$\vn p+\vn{q}+\vn{k}$}{v1,o1}
\fmf{plain_arrow, l.side=right, label=$\vn p$}{i1,v1}
\fmf{photon, label=$\vn q$}{i2,v1}
\fmf{photon, l.side=left, label=$\vn k$}{i3,v1}
\fmfv{d.shape=square, d.size=3thick}{v1}
\end{fmfgraph*}
}&=-V(\vn p,-\vn q)
\end{align*}
\end{fmffile}
\caption{Diagrammatic notation}
\label{fig:notazione}
\end{figure}
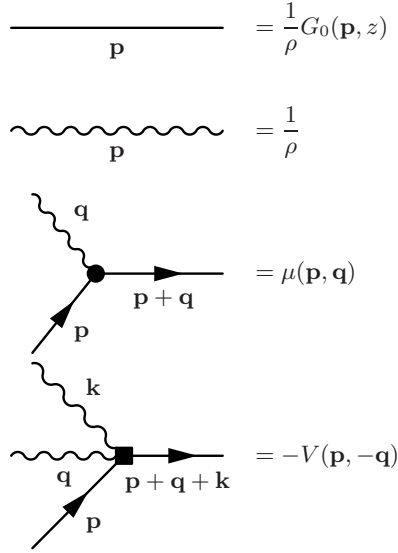 

Fig.~\ref{fig:notazione} defines our diagrammatic notation.
Note that the vertex $V_4$ is not symmetric under the interchange of
the two $\delta\rho$ lines. Now we are able to write down the one-loop
diagrams.

Recalling that the resolvent $G(\vn p,z)$ is given by
Eq.~(\ref{eq:G_deltarho}), we compute the one-loop contribution to
$\overline{\left<\phi(\vn{x})\phi(\vn{y})\right>}$:
\begin{fmffile}{L_1_1}
\begin{align}
L_1^{(1)} \equiv \parbox{35mm}{
\begin{fmfgraph*}(100,50)
\fmfleft{i1}
\fmfright{o1}
\fmf{plain, tension=2}{i1,v1}
\fmf{photon, tension=.3,left}{v1,v2}
\fmf{plain, tension=1}{v1,v2}
\fmf{plain, tension=2}{v2,o1}
\fmfv{d.shape=circle, d.size=3thick}{v1,v2}
\end{fmfgraph*}
}&=\frac{G_0^2(\vn p,z)}{\rho}\int \frac{\mathrm{d}^D\vn q}{(2\pi)^D}G_0(\vn p-\vn q,z)\mu^2(\vn p,-\vn q)\\
\parbox{35mm}{
\begin{fmfgraph*}(100,40)
\fmfleft{i1}
\fmfright{o1}
\fmf{plain}{i1,v1}
\fmf{photon, tension=1.2,right}{v1,v1}
\fmf{plain}{v1,o1}
\fmfv{d.shape=square, d.size=3thick}{v1}
\end{fmfgraph*}
}&=0
\end{align}
\end{fmffile}
The last diagram gives a general result: every tadpole made with a
vertex with four fields gives a vanishing contribution due to the form
of the vertex. The term with one external $\delta\rho$ insertion,
arising from
$\overline{\delta\rho(\vn{x})\left<\phi(\vn{x})\phi(\vn{y})\right>}$,
is given by
\begin{fmffile}{L_1_2}
\begin{equation}
L_2^{(1)} \equiv \parbox{29mm}{
\begin{fmfgraph*}(70,50)
\fmfleft{i1}
\fmfright{o1}
\fmf{plain}{i1,v1}
\fmf{plain,tension=1.5}{v1,o1}
\fmf{photon, tension=.5, left}{i1,v1}
\fmfv{d.shape=circle, d.size=3thick}{v1}
\end{fmfgraph*}
}=-\frac{2G^2_0(\vn p,z)}{\rho}\int \frac{\mathrm{d}^D\vn q}{(2\pi)^D}G_0(\vn p-\vn q,z)\mu(\vn p,-\vn q)
\end{equation}
\end{fmffile}
The last contribution to the self-energy at one loop comes from
$\overline{\delta\rho(\vn{x})\delta\rho(\vn{y})\left<\phi(\vn{x})\phi(\vn{y})\right>}$,
and is
\begin{fmffile} {L_1_3}
\begin{equation}
L_3^{(1)} \equiv \parbox{18mm}{
\begin{fmfgraph*}(50,40)
\fmfleft{i1}
\fmfright{o1}
\fmf{plain}{i1,o1}
\fmf{photon, tension=.8, left}{i1,o1}
\end{fmfgraph*}
}=\frac{1}{\rho}\int \frac{\mathrm{d}^D\vn q}{(2\pi)^D}G_0(\vn p-\vn q,z)
\end{equation}
\end{fmffile}

Note that this last contribution has an ultraviolet divergence since
the propagator goes to a finite costant when the internal momentum
goes to infinity. Nevertheless, by adding the four diagrams
\begin{align}
L_1^{(1)}+L_2^{(1)}+L_1^{(3)}&=\frac{1}{\rho} \int
\frac{\mathrm{d}^D\vn q}{(2\pi)^D}G_0(\vn p-\vn q,z)\left[\mu(\vn p,-\vn q)G_0(\vn p,z)-1
\right]^2=\\ 
&=\frac{G_0^2(\vn p,z)}{\rho}\int \frac{\mathrm{d}^D\vn q}{(2\pi)^D}G_0(\vn q,z)V^2(\vn q,\vn p)\equiv G_0^2(\vn p,z)\Sigma^{(1)}(\vn p,z),
\end{align}
the divergence disappears and one recovers the combinatorial result for
the one-loop self-energy.

\subsection{Two loops}

Let us first consider the two-loop diagrams
arising from $\overline{\left<\phi(\vn{x})\phi(\vn{y})\right>}$:
\begin{fmffile}{L_2_1}
\begin{align}
L_1^{(2)} \equiv \parbox{27mm}{
\begin{fmfgraph*}(80,40)
\fmfleft{i1}
\fmfright{o1}
\fmf{plain, tension=4}{i1,v1}
\fmf{plain, tension=2}{v1,v2}
\fmf{plain, tension=.7}{v2,v3}
\fmf{plain, tension=2}{v3,v4}
\fmf{plain, tension=4}{v4,o1}
\fmf{photon, tension=.7, left}{v1,v4}
\fmf{photon, tension=.7, left}{v2,v3}
\fmfv{d.shape=circle, d.size=3thick}{v1,v2,v3,v4}
\end{fmfgraph*}
}&\,=\frac{G_0^2(\vn p,z)}{\rho^2}\int \frac{\mathrm{d}^D\vn q\mathrm{d}^D\vn k}{(2\pi)^{2D}}G_0^2(\vn q,z)G_0(\vn{q}-\vn{k},z)\mu^2(\vn p,\vn q-\vn p)\mu^2(\vn q,-\vn k)\\
L_2^{(2)} \equiv \parbox{27mm}{
\begin{fmfgraph*}(80,40)
\fmfleft{i1}
\fmfright{o1}
\fmf{plain, tension=4}{i1,v1}
\fmf{plain, tension=2}{v1,v2}
\fmf{plain, tension=.7}{v2,v3}
\fmf{plain, tension=2}{v3,v4}
\fmf{plain, tension=4}{v4,o1}
\fmf{photon, tension=.7, left}{v1,v3}
\fmf{photon, tension=.7, left}{v2,v4}
\fmfv{d.shape=circle, d.size=3thick}{v1,v2,v3,v4}
\end{fmfgraph*}
}&\,=\frac{G_0^2(\vn p,z)}{\rho^2}\int \frac{\mathrm{d}^D\vn q\mathrm{d}^D\vn k}{(2\pi)^{2D}}G_0(\vn q,z)G_0(\vn k,z)G_0(\vn p-\vn{q}-\vn{k},z)\mu(\vn p,\vn q-\vn p)\mu(\vn p,\vn k-\vn p)\cdot\\
&\quad \cdot \ \mu(\vn q,\vn k-\vn p)\mu(\vn k,\vn q-\vn p)\\
L_3^{(2)} \equiv \parbox{27mm}{
\begin{fmfgraph*}(80,40)
\fmfleft{i1}
\fmfright{o1}
\fmf{plain, tension=4}{i1,v1}
\fmf{plain, tension=.7}{v1,v2}
\fmf{plain, tension=.7}{v2,v3}
\fmf{plain, tension=4}{v3,o1}
\fmf{photon, tension=.7, left}{v1,v2}
\fmf{photon, tension=.7, left}{v2,v3}
\fmfv{d.shape=circle, d.size=3thick}{v1,v2,v3}
\end{fmfgraph*}
}&\,=-\frac{G_0^2(\vn p,z)}{\rho^2}\int \frac{\mathrm{d}^D\vn q\mathrm{d}^D\vn k}{(2\pi)^{2D}}G_0(\vn q,z)G_0(\vn k,z)\mu(\vn p,\vn q-\vn p)\mu(\vn p,\vn k-\vn q)\mu(\vn p,\vn k-\vn p)\\
L_4^{(2)} \equiv \parbox{27mm}{
\begin{fmfgraph*}(80,40)
\fmfleft{i1}
\fmfright{o1}
\fmf{plain, tension=4}{i1,v1}
\fmf{plain, tension=.7}{v1,v2}
\fmf{plain, tension=.7}{v2,v3}
\fmf{plain, tension=4}{v3,o1}
\fmf{photon, tension=.7, left}{v1,v2}
\fmf{photon, tension=.7, left}{v2,v3}
\fmfv{d.shape=circle, d.size=3thick}{v1,v3}
\fmfv{d.shape=square, d.size=3thick}{v2}
\end{fmfgraph*}
}&\,=\frac{G_0^2(\vn p,z)}{\rho^2}\int \frac{\mathrm{d}^D\vn q\mathrm{d}^D\vn k}{(2\pi)^{2D}}G_0(\vn q,z)G_0(\vn k,z)\mu(\vn p,\vn q-\vn p)\mu(\vn p,\vn k-\vn p)\cdot\\
&\quad \cdot \left[V(\vn p-\vn q,-\vn q)-V(\vn k-\vn p,-\vn q)\right]\\
L_5^{(2)} \equiv \parbox{27mm}{
\begin{fmfgraph*}(80,40)
\fmfleft{i1}
\fmfright{o1}
\fmf{plain, tension=4}{i1,v1}
\fmf{plain, tension=2}{v1,v2}
\fmf{plain, tension=.7}{v2,v3}
\fmf{plain, tension=4}{v3,o1}
\fmf{photon, tension=.7, left}{v1,v3}
\fmf{photon, tension=.7, left}{v2,v3}
\fmfv{d.shape=circle, d.size=3thick}{v1,v2}
\fmfv{d.shape=square, d.size=3thick}{v3}
\end{fmfgraph*}
}&\,=-\frac{2G_0^2(\vn p,z)}{\rho^2} \int \frac{\mathrm{d}^D\vn q\mathrm{d}^D\vn k}{(2\pi)^{2D}}G_0(\vn q,z)G_0(\vn{q}-\vn{k},z)\mu(\vn q,-\vn k)\cdot\\
&\quad\cdot \left[-V(\vn p-\vn q,\vn k-\vn q)-V(\vn k,\vn k-\vn q)\right]\\
L_6^{(2)} \equiv \parbox{27mm}{
\begin{fmfgraph*}(80,40)
\fmfleft{i1}
\fmfright{o1}
\fmf{plain, tension=4}{i1,v1}
\fmf{plain, tension=.7}{v1,v2}
\fmf{plain, tension=4}{v2,o1}
\fmf{photon, tension=.7, left}{v1,v2}
\fmf{photon, tension=.7, right}{v1,v2}
\fmfv{d.shape=square, d.size=3thick}{v1,v2}
\end{fmfgraph*}
}&\,=\frac{G_0^2(\vn p,z)}{\rho^2} \int \frac{\mathrm{d}^D\vn q\mathrm{d}^D\vn k}{(2\pi)^{2D}}G_0(\vn p-\vn{q}-\vn{k},z)V(\vn q,\vn p)\left[V(\vn q,\vn k+\vn q-\vn p)+V(\vn k,\vn k+\vn q-\vn p)\right]\\
L_7^{(2)} \equiv \parbox{27mm}{
\begin{fmfgraph*}(80,40)
\fmfleft{i1}
\fmfright{o1}
\fmf{plain, tension=4}{i1,v1}
\fmf{plain, tension=2}{v1,v2}
\fmf{plain, tension=4}{v2,v3}
\fmf{plain, tension=2}{v3,v4}
\fmf{plain, tension=4}{v4,o1}
\fmf{photon, tension=.7, left}{v1,v2}
\fmf{photon, tension=.7, left}{v3,v4}
\fmfv{d.shape=circle, d.size=3thick}{v1,v2,v3,v4}
\end{fmfgraph*}
}&\,=\frac{G_0^3(\vn p,z)}{\rho^2} \int \frac{\mathrm{d}^D\vn q\mathrm{d}^D\vn k}{(2\pi)^{2D}} G_0(\vn k,z)G_0(\vn q,z) \mu^2(\vn p,\vn q-\vn p)\mu^2(\vn p,\vn k-\vn p)
\end{align}
\end{fmffile}

The diagram $L_7^{(2)}$ seems to be already included in the Dyson
resummation of the one-loop result. However, since diagrams with one
and zero external legs have to be included in the diagrammatic
expansion, it also provides a genuine contribution to the two-loop
result.  Note that in order to obatin $L_3^{(2)}$ one uses
Eq.~(\ref{eq:delta-rho-3-points}) for the $3$-point correlation of
$\delta\rho$. The other diagrams involve only the disconnected part of
the $4$-point correlation function, while the connected one would only
matter at three loops.

Next we must consider the contribution arising from
$\overline{\delta\rho(\vn{x})\left<\phi(\vn{x})\phi(\vn{y})\right>}$:
\begin{fmffile}{L_2_2}
\begin{align}
L_8^{(2)} \equiv \ \parbox{27mm}{
\begin{fmfgraph*}(100,40)
\fmfleft{i1}
\fmfright{o1}
\fmf{plain, tension=2}{i1,v1}
\fmf{plain, tension=.7}{v1,v2}
\fmf{plain, tension=2}{v2,v3}
\fmf{plain, tension=4}{v3,o1}
\fmf{photon, tension=.7, left}{i1,v3}
\fmf{photon, tension=.7, left}{v1,v2}
\fmfv{d.shape=circle, d.size=3thick}{v1,v2,v3}
\end{fmfgraph*}
}&\,=-\frac{2G_0(\vn p,z)}{\rho^2}\int \frac{\mathrm{d}^D\vn q\mathrm{d}^D\vn k}{(2\pi)^{2D}}G_0^2(\vn q,z)G_0(\vn{q}-\vn{k},z)\mu(\vn p,\vn q-\vn p)\mu^2(\vn q,-\vn k)\\
L_9^{(2)} \equiv \ \parbox{27mm}{
\begin{fmfgraph*}(80,40)
\fmfleft{i1}
\fmfright{o1}
\fmf{plain, tension=2}{i1,v1}
\fmf{plain, tension=.7}{v1,v2}
\fmf{plain, tension=2}{v2,v3}
\fmf{plain, tension=4}{v3,o1}
\fmf{photon, tension=.7, left}{i1,v2}
\fmf{photon, tension=.7, left}{v1,v3}
\fmfv{d.shape=circle, d.size=3thick}{v1,v2,v3}
\end{fmfgraph*}
}&\,=-\frac{2G_0(\vn p,z)}{\rho^2}\int \frac{\mathrm{d}^D\vn q\mathrm{d}^D\vn k}{(2\pi)^{2D}}G_0(\vn q,z)G_0(\vn k,z)G_0(\vn p-\vn{q}-\vn{k},z)\mu^2(\vn k,\vn q-\vn p)\cdot\\
&\quad\cdot\ \mu(\vn q,\vn k-\vn p)\mu(\vn p,\vn k-\vn p)
\end{align}
\begin{align}
L_{10}^{(2)} \equiv \ \parbox{27mm}{
\begin{fmfgraph*}(80,40)
\fmfleft{i1}
\fmfright{o1}
\fmf{plain}{i1,v1}
\fmf{plain}{v1,v2}
\fmf{plain, tension=4}{v2,o1}
\fmf{photon, tension=.7, left}{i1,v1}
\fmf{photon, tension=.7, left}{v1,v2}
\fmfv{d.shape=circle, d.size=3thick}{v1,v2}
\end{fmfgraph*}
}&\,=\frac{2G_0(\vn p,z)}{\rho^2}\int \frac{\mathrm{d}^D\vn q\mathrm{d}^D\vn k}{(2\pi)^{2D}}G_0(\vn q,z)G_0(\vn k,z)\mu(\vn q,\vn k-\vn q)\mu(\vn p,\vn k-\vn p)\\
L_{11}^{(2)} \equiv \ \parbox{27mm}{
\begin{fmfgraph*}(80,40)
\fmfleft{i1}
\fmfright{o1}
\fmf{plain, tension=2}{i1,v1}
\fmf{plain}{v1,v2}
\fmf{plain, tension=4}{v2,o1}
\fmf{photon, tension=.7, left}{i1,v2}
\fmf{photon, tension=.7, left}{v1,v2}
\fmfv{d.shape=circle, d.size=3thick}{v1}
\fmfv{d.shape=square, d.size=3thick}{v2}
\end{fmfgraph*}
}&\,=-\frac{2G_0(\vn p,z)}{\rho^2}\int \frac{\mathrm{d}^D\vn q\mathrm{d}^D\vn k}{(2\pi)^{2D}}G_0(\vn q,z)G_0(\vn k,z)\mu(\vn q,\vn k-\vn q)\left[V(\vn p-\vn q,-\vn k)+V(\vn{q}-\vn{k},-\vn k)\right]\\
L_{12}^{(2)} \equiv \ \parbox{27mm}{
\begin{fmfgraph*}(80,40)
\fmfleft{i1}
\fmfright{o1}
\fmf{plain}{i1,v1}
\fmf{plain}{v1,v2}
\fmf{plain, tension=4}{v2,o1}
\fmf{photon, tension=.7, left}{i1,v1}
\fmf{photon, tension=.7, left}{v1,v2}
\fmfv{d.shape=square, d.size=3thick}{v1}
\fmfv{d.shape=circle, d.size=3thick}{v2}
\end{fmfgraph*}
}&\,=-\frac{2G_0(\vn p,z)}{\rho^2}\int \frac{\mathrm{d}^D\vn q\mathrm{d}^D\vn k}{(2\pi)^{2D}}G_0(\vn q,z)G_0(\vn k,z)\mu(\vn p,\vn k-\vn p)\left[V(\vn p-\vn q,-\vn q)+V(\vn k-\vn p,-\vn q)\right]\\
L_{13}^{(2)} \equiv \ \parbox{27mm}{
\begin{fmfgraph*}(80,40)
\fmfleft{i1}
\fmfright{o1}
\fmf{plain}{i1,v1}
\fmf{plain, tension=4}{v1,v2}
\fmf{plain}{v2,v3}
\fmf{plain, tension=4}{v3,o1}
\fmf{photon, tension=.7, left}{i1,v1}
\fmf{photon, tension=.7, left}{v2,v3}
\fmfv{d.shape=circle, d.size=3thick}{v1,v2,v3}
\end{fmfgraph*}
}&\,=-\frac{2G_0^2(\vn p,z)}{\rho^2} \int \frac{\mathrm{d}^D\vn q\mathrm{d}^D\vn k}{(2\pi)^{2D}}G_0(\vn q,z)G_0(\vn k,z)\mu^2(\vn p,\vn k-\vn p)\mu(\vn p,\vn q-\vn p)
\end{align}
\end{fmffile}
As before, we have used the disconnected part of the $4$-point
function, appart from $L_{10}^{(2)}$ where we have used the $3$-point
function. Note also that $L_{13}^{(2)}$ arises both in the Dyson
resummation and in the two-loop expansion.

Finally, we consider the diagrams
arising from
$\overline{\delta\rho(\vn{x})\delta\rho(\vn{y})\left<\phi(\vn{x})\phi(\vn{y})\right>}$:
\begin{fmffile} {L_2_3}
\begin{align}
L_{14}^{(2)} \equiv \ \parbox{23mm}{
\begin{fmfgraph*}(60,40)
\fmfleft{i1}
\fmfright{o1}
\fmf{plain, tension=2}{i1,v1}
\fmf{plain}{v1,v2}
\fmf{plain, tension=2}{v2,o1}
\fmf{photon, tension=.9, left}{i1,o1}
\fmf{photon, tension=.9, left}{v1,v2}
\fmfv{d.shape=circle, d.size=3thick}{v1,v2}
\end{fmfgraph*}
}&=\frac{1}{\rho^2} \int \frac{\mathrm{d}^D\vn q\mathrm{d}^D\vn k}{(2\pi)^{2D}}G_0^2(\vn q,z)G_0(\vn{q}-\vn{k},z)\mu^2(\vn q,-\vn k)\\
L_{15}^{(2)} \equiv \ \parbox{23mm}{
\begin{fmfgraph*}(60,40)
\fmfleft{i1}
\fmfright{o1}
\fmf{plain, tension=2}{i1,v1}
\fmf{plain}{v1,v2}
\fmf{plain, tension=2}{v2,o1}
\fmf{photon, tension=.9, left}{i1,v2}
\fmf{photon, tension=.9, left}{v1,o1}
\fmfv{d.shape=circle, d.size=3thick}{v1,v2}
\end{fmfgraph*}
}&=\frac{1}{\rho^2} \int \frac{\mathrm{d}^D\vn q\mathrm{d}^D\vn k}{(2\pi)^{2D}}G_0(\vn q,z)G_0(\vn k,z)G_0(\vn p-\vn{q}-\vn{k},z)\mu^2(\vn q,\vn k-\vn p)\mu(\vn k,\vn q-\vn p)\\
L_{16}^{(2)} \equiv \ \parbox{23mm}{
\begin{fmfgraph*}(60,40)
\fmfleft{i1}
\fmfright{o1}
\fmf{plain}{i1,v1}
\fmf{plain}{v1,o1}
\fmf{photon, tension=.9, left}{i1,v1}
\fmf{photon, tension=.9, left}{v1,o1}
\fmfv{d.shape=circle, d.size=3thick}{v1}
\end{fmfgraph*}
}&=-\frac{1}{\rho^2} \int \frac{\mathrm{d}^D\vn q\mathrm{d}^D\vn k}{(2\pi)^{2D}}G_0(\vn q,z)G_0(\vn k,z)\mu(\vn q,\vn k-\vn q)\\
L_{17}^{(2)} \equiv \ \parbox{23mm}{
\begin{fmfgraph*}(60,40)
\fmfleft{i1}
\fmfright{o1}
\fmf{plain}{i1,v1}
\fmf{plain}{v1,o1}
\fmf{photon, tension=.9, left}{i1,v1}
\fmf{photon, tension=.9, left}{v1,o1}
\fmfv{d.shape=square, d.size=3thick}{v1}
\end{fmfgraph*}
}&=\frac{1}{\rho^2} \int \frac{\mathrm{d}^D\vn q\mathrm{d}^D\vn k}{(2\pi)^{2D}}G_0(\vn q,z)G_0(\vn k,z)\left[V(\vn p-\vn q,-\vn q)+V(\vn k-\vn p,-\vn q)\right]\\
L_{18}^{(2)} \equiv \ \parbox{23mm}{
\begin{fmfgraph*}(60,40)
\fmfleft{i1}
\fmfright{o1}
\fmf{plain}{i1,v1}
\fmf{plain, tension=2}{v1,v2}
\fmf{plain}{v2,o1}
\fmf{photon, tension=.9, left}{i1,v1}
\fmf{photon, tension=.9, left}{v2,o1}
\fmfv{d.shape=circle, d.size=3thick}{v1,v2}
\end{fmfgraph*}
}&=\frac{1}{\rho^2} \int \frac{\mathrm{d}^D\vn q\mathrm{d}^D\vn k}{(2\pi)^{2D}}G_0(\vn q,z)G_0(\vn k,z) \mu(\vn p,\vn q-\vn p)\mu(\vn p,\vn k-\vn p)
\end{align}
\end{fmffile}

We now show how the diagrams can be summed up to give the
combinatorial expressions for the self energy. Consider the diagram
$L_1^{(2)}$; it has the same topology
(in the sense of momenta flow and vertex positions) of
$\Sigma_A^{(2)}$. In fact, it can be combined with $L_8^{(2)}$ and
$L_{14}^{(2)}$ to give:
\begin{align}
L_1^{(2)}+L_8^{(2)}+L_{14}^{(2)}&=\frac{1}{\rho^2} \int \frac{\mathrm{d}^D\vn q\mathrm{d}^D\vn k}{(2\pi)^{2D}}G_0^2(\vn p,z)G_0^2(\vn q,z)G_0(\vn{q}-\vn{k},z)V^2(\vn p-\vn q,\vn p)\mu^2(\vn q,-\vn k)=\\
&=\Sigma_A^{(2)}(\vn p,z)+\Omega_1(\vn p,z),
\end{align}
where we have defined
\begin{equation}
\Omega_1(\vn p,z)=\frac{1}{\rho^2} \int \frac{\mathrm{d}^D\vn q\mathrm{d}^D\vn k}{(2\pi)^{2D}}\left[G_0^2(\vn p,z)G_0(\vn k,z)V^2(\vn p-\vn q,\vn p)-2G_0^2(\vn p,z)G_0(\vn q,z)G_0(\vn k,z)V^2(\vn p-\vn q,\vn p)V(\vn{q}-\vn{k},q)\right].
\end{equation}
In the same way we can combine $L_2^{(2)}$, $L_9^{(2)}$ and
$L_{15}^{(2)}$; they have the same topology of $\Sigma_B(\vn p,z)$:
\begin{align}
L_2^{(2)}+L_9^{(2)}+L_{15}^{(2)} &= \frac{1}{\rho^2} \int
\frac{\mathrm{d}^D\vn q\mathrm{d}^D\vn k}{(2\pi)^{2D}}G_0(\vn q,z)G_0(\vn k,z)G_0(\vn p-\vn{q}-\vn{k},z)\mu(\vn q,\vn k-\vn p)\mu(\vn k,\vn q-\vn p)\cdot
\\
&\quad \cdot V(\vn p-\vn q,\vn p)V(\vn p-\vn k,\vn p)= \Sigma_B^{(2)}(\vn p,z)+\Omega_2(\vn p,z)
\end{align}
where
\begin{align}
\Omega_2(\vn p,z)&=\frac{1}{\rho^2} \int \frac{\mathrm{d}^D\vn q\mathrm{d}^D\vn k}{(2\pi)^{2D}}\left[G_0^2(\vn p,z)G_0(\vn p-\vn{q}-\vn{k},z)V(\vn p-\vn q,\vn p)V(\vn p-\vn k,\vn p)+\right. \\
&\quad \left. -2G_0^2(\vn p,z)G_0(\vn q,z)G_0(\vn k,z)
  V(\vn{q}+\vn{k},\vn p)V(\vn p-\vn k,\vn p)V(\vn{q}+\vn{k},\vn k)\right]. 
\end{align}

We now add the diagrams $L_7^{(2)}$, $L_{13}^{(2)}$ and $L_{18}^{(2)}$
because they produce the Dyson resummation of the self energy at one
loop that we want to isolate from the other contributions that have to
be included in the self energy at two loops. They give
\begin{align}
L_7^{(2)}+L_{13}^{(2)}+L_{18}^{(2)}&=\frac{1}{\rho^2} \int \frac{\mathrm{d}^D\vn q\mathrm{d}^D\vn k}{(2\pi)^{2D}}G_0^3(\vn p,z)G_0(\vn q,z)G_0(\vn k,z)V(\vn p-\vn q,\vn p)V(\vn p-\vn k,\vn p)\mu(\vn p,\vn q-\vn p)\\
&\quad \cdot \mu(\vn p,\vn k-\vn p)= G_0(\vn p,z)\Sigma^{(1)}(\vn p,z)G_0(\vn p,z)\Sigma^{(1)}(\vn p,z)G_0(\vn p,z)+\Omega_3(\vn p,z),
\end{align}
where 
\begin{equation}
\Omega_3(\vn p,z)= \frac{1}{\rho^2} \int \frac{\mathrm{d}^D\vn q\mathrm{d}^D\vn k}{(2\pi)^{2D}}\left[G_0(\vn p,z)G_0(\vn q,z)G_0(\vn k,z)V(\vn p-\vn q,\vn p)V(\vn p-\vn k,\vn p)\left( 1-2G_0(\vn p,z)V(\vn p-\vn q,\vn p)\right) \right].
\end{equation} 
At this point one can check that
\begin{equation}
\Omega_1(\vn p,z)+\Omega_2(\vn p,z)+\Omega_3(\vn p,z)+L_3^{(2)}+L_4^{(2)}+L_5^{(2)}+L_6^{(2)}+L_{10}^{(2)}+L_{11}^{(2)}+L_{12}^{(2)}+L_{16}^{(2)}+L_{17}^{(2)}=\Sigma_C^{(2)}(\vn p,z)
\end{equation}
and the combinatorial result is recovered.

\subsection{The small $p$ behaviour}

We will now prove that the prefactor of the term $p^2
\lambda^{(D-2)/2}$ is zero to all orders in perturbation theory. For
this purpose, the field theory approach turns out to be very
convenient.

Consider the vertex $V_3$ with three fields,
\begin{fmffile}{diag}
\begin{equation}
\parbox{25mm}{
\begin{fmfgraph*}(60,40)
\fmfleft{i1,i2}
\fmfright{o1}
\fmf{plain, label=$p$}{i1,v1}
\fmf{plain, label=$q$}{i2,v1}
\fmf{photon, label=$p+q$}{v1,o1}
\fmfv{d.shape=circle, d.size=3thick}{v1}
\end{fmfgraph*}
}=G_0^{-1}(\vn p,z)+V(\vn q,\vn p)=z-\rho\hat f(\vn 0)+\rho \hat f(\vn p)+\rho \hat f (\vn q)-\rho\hat f(\vn p-\vn q).
\end{equation}
\end{fmffile}
It is easy to see that this vertex is symmetric and can be written as
\begin{equation}
V_3=z+S(\vn p,\vn q), \qquad \text{where}\qquad S(\vn p,\vn q)=S(\vn q,\vn p).
\end{equation} 
Moreover one can check directly that
\begin{equation}
S(\vn p,\vn 0)=S(\vn 0,\vn p)=0.
\end{equation} 
Consider now the vertex with four fields. We see that the Wick
contractions between the fields $\delta\rho$ symmetrize the vertex. In
fact in every diagram this vertex appears in the form
\begin{fmffile}{diag2}
\begin{align}
\parbox{28mm}{
\begin{fmfgraph*}(80,40)
\fmfbottom{i1,i2}
\fmftop{o1,o2}
\fmf{fermion, label=$p$, l.side=right}{i1,v1}
\fmf{fermion, label=$q$, l.side=right}{v1,i2}
\fmf{photon, label=$k$, l.side=right}{v1,o1}
\fmf{photon, label=$q-p-k$, l.side=left}{v1,o2}
\fmfv{d.shape=square, d.size=3thick}{v1}
\end{fmfgraph*}
}&=-\left(V(\vn k,-\vn p)+V(\vn q-\vn p-\vn k,-\vn p)\right)\\
 &= -\rho\left(\hat f(\vn k)-\hat f(\vn p+\vn k)+\hat f(\vn q-\vn p-\vn k)-\hat f (\vn q-\vn k)\right).
\end{align}
\end{fmffile}The important thing is that this vertex vanishes when one of the two $G_0$
bare propagators carries a null momentum. Consider now a diagram that
arises from the expansion of the resolvent $G(p,z)$. At the lowest
order in $z$ when the diagram contains some three-field vertices one
has to consider only the symmetric part of these vertices.  Let us
apply the method explained above in order to extract the contribution
to the self energy proportional to $z^{(D-2)/2}$. Apparently, if one
sets to zero the momentum of a bare propagator that enters into a
vertex then its contribution to the imaginary part vanishes. This
seems very strange because from this argument it follows that only
$L_3^{(1)}$ contributes to the imaginary part. Moreover if we consider
the two-loop contributions we see that there are no contributions to
the imaginary part of the self energy because the diagrammatic
expansion $L_1^{(2)}$-$L_{18}^{(2)}$ contains at least one vertex that
vanishes when we set to zero one of the momentum brought by a
$\phi$-propagator.  However the argument is not complete. Actually,
the diagrammatic expansion $L_1^{(2)}$-$L_{18}^{(2)}$ does not
contain only the contribution
\begin{equation}
G_0(\vn p,z)\Sigma^{(2)}(\vn p,z)G_0(\vn p,z)
\end{equation}
since it contains also the Dyson resummation of
the one loop self energy. This is the fact that completes the argument
and will lead us to prove that a contribution proportional to
$z^{(D-2)/2}p^2$ cannot appear at any order in perturbation
theory.

We start checking the argument just given at the one-loop
level.  Let us introduce the notation
\begin{fmffile}{diag3}
\begin{equation}
\parbox{20mm}{
\begin{fmfgraph*}(50,30)
\fmfleft{i1}
\fmfright{o1}
\fmf{dashes}{i1,o1}
\end{fmfgraph*}
}=G_0^{-1}(\vn p,z),
\end{equation}
\end{fmffile}
so that the self energy at one loop can be written
diagrammatically as
\begin{fmffile}{diag4}
\begin{equation}
\Sigma^{(1)}(\vn p,z)=
\ \ \parbox{20mm}{ 
\begin{fmfgraph*}(50,50)
\fmfleft{i1}
\fmfright{o1}
\fmf{plain}{i1,o1}
\fmf{photon, tension=.7, left}{i1,o1}
\fmfv{d.shape=circle, d.size=3thick}{i1,o1}
\end{fmfgraph*}
}+\ \ \parbox[c]{27mm}{
\begin{fmfgraph*}(70,50)
\fmfleft{i1}
\fmfright{o1}
\fmf{dashes, tension=2}{i1,v1}
\fmf{plain}{v1,o1}
\fmf{photon, tension=.7, left}{v1,o1}
\fmfv{d.shape=circle, d.size=3thick}{v1}
\end{fmfgraph*}
}+\ \ \parbox[c]{32mm}{
\begin{fmfgraph*}(70,50)
\fmfleft{i1}
\fmfright{o1}
\fmf{dashes, tension=2}{i1,v1}
\fmf{dashes, tension=2}{v2,o1}
\fmf{plain}{v1,v2}
\fmf{photon, tension=.7, left}{v1,v2}
\end{fmfgraph*}
}.
\end{equation}
\end{fmffile}
From this expansion and from the above argument one sees that the
imaginary part of the self energy (we will refer always to the
imaginary part proportional to $z^{(D-2)/2}$) may come from the last
diagram and is correctly given by
\begin{equation}
\tilde{\text{Im}} \lim_{\epsilon\to 0^+}\Sigma^{(1)}(\vn
p,z+i\epsilon)\propto z^{(D-2)/2}V^2(\vn p,\vn p), 
\end{equation} 
which is also the contribution that can be easily calculated from the
combinatorial expression. Now consider the expansion at two loops.
From the combinatorial expressions of the self energy we
immediately see that the immaginary part comes from only
$\Sigma_C^{(2)}$ and can be rewritten in the form
\begin{equation}
\tilde{\text{Im}} \lim_{\epsilon\to 0} \Sigma^{(2)}(\vn
p,z+i\epsilon)\propto -2z^{(D-2)/2} V(\vn p,\vn p)\Sigma^{(1)}(\vn p,z). 
\end{equation} 
Consider now the diagrammatic expansion for the two loop self-energy
$L_1^{(2)}$-$L_{18}^{(2)}$. We have to extract from this expansion the
term
\begin{equation}
\Sigma^{(1)}(\vn p,z)G_0(\vn p,z)\Sigma^{(1)}(\vn p,z).
\label{eq:GSGSG}
\end{equation} 
Now we will do this in a diagrammatic way. Consider the diagrammatic
expansion for the above term (Fig.~\ref{fig:diag_5}).

\begin{figure}[h]
\centering
\begin{fmffile}{diag5}
\begin{equation}
\Sigma^{(1)}(\vn p,z)\risz \Sigma^{(1)}(\vn p,z)=
\end{equation}
\begin{equation}
= \ \ \biggl\{\ 
\parbox[b][1cm][b]{11mm}{
\begin{fmfgraph*}(30,30)
\fmfbottom{i1,o1}
\fmf{plain}{i1,o1}
\fmf{photon, tension=.8, left}{i1,o1}
\fmfv{d.shape=circle, d.size=3thick}{i1,o1}
\end{fmfgraph*}
}\ \ +\ \ 
\parbox[b][1cm][b]{30mm}{
\begin{fmfgraph*}(70,50)
\fmfbottom{i1,o1}
\fmf{dashes, tension=2}{i1,v1}
\fmf{plain}{v1,o1}
\fmf{photon, tension=.8, left}{v1,o1}
\fmfv{d.shape=circle, d.size=3thick}{o1}
\end{fmfgraph*}
}\ +\ \ 
\parbox[b][1cm][b]{30mm}{
\begin{fmfgraph*}(70,50)
\fmfbottom{i1,o1}
\fmf{dashes, tension=2}{i1,v1}
\fmf{dashes, tension=2}{v2,o1}
\fmf{plain}{v1,v2}
\fmf{photon, tension=.8, left}{v1,v2}
\end{fmfgraph*}
}\biggr\} \cdot
\end{equation}
\begin{equation}
\cdot\ \  
\parbox[b][1cm][b]{20mm}{
\begin{fmfgraph*}(50,50)
\fmfbottom{i1,o1}
\fmf{plain}{i1,o1}
\end{fmfgraph*}
}\ \cdot \biggl\{\  
\parbox[b][1cm][b]{25mm}{
\begin{fmfgraph*}(30,30)
\fmfbottom{i1,o1}
\fmf{plain}{i1,o1}
\fmf{photon, tension=.8, left}{i1,o1}
\fmfv{d.shape=circle, d.size=3thick}{i1,o1}
\end{fmfgraph*}
}\ \ +\ \ 
\parbox[b][1cm][b]{30mm}{
\begin{fmfgraph*}(70,50)
\fmfbottom{i1,o1}
\fmf{dashes, tension=2}{i1,v1}
\fmf{plain}{v1,o1}
\fmf{photon, tension=.8, left}{v1,o1}
\fmfv{d.shape=circle, d.size=3thick}{o1}
\end{fmfgraph*}
}\ +\ \ 
\parbox[b][1cm][b]{30mm}{
\begin{fmfgraph*}(70,50)
\fmfbottom{i1,o1}
\fmf{dashes, tension=2}{i1,v1}
\fmf{dashes, tension=2}{v2,o1}
\fmf{plain}{v1,v2}
\fmf{photon, tension=.8, left}{v1,v2}
\end{fmfgraph*}
}\ \biggr\}=
\end{equation}
\begin{gather}
=2\Sigma^{(1)}(\vn p,\,z)\cdot\ 
\parbox[b][1cm][b]{30mm}{
\begin{fmfgraph*}(70,50)
\fmfbottom{i1,o1}
\fmf{plain}{i1,v1}
\fmf{dashes, tension=2}{v1,o1}
\fmf{photon, tension=.5, left}{i1,v1}
\end{fmfgraph*}
}\ - \ \
\parbox[b][1cm][b]{30mm}{
\begin{fmfgraph*}(60,30)
\fmfbottom{i1,o1}
\fmf{dashes, tension=2}{i1,v1}
\fmf{plain}{v1,v2}
\fmf{dashes, tension=2}{v2,v3}
\fmf{plain}{v3,v4}
\fmf{dashes, tensio=2}{v4,o1}
\fmf{photon, tension=.5, left}{v1,v2}
\fmf{photon, tension=.5, left}{v3,v4}
\end{fmfgraph*}
}+\\
+\ \,	
\parbox[b][1cm][b]{30mm}{
\begin{fmfgraph*}(70,50)
\fmfbottom{v1,v4}
\fmf{plain}{v1,v2}
\fmf{plain, tension=2}{v2,v3}
\fmf{plain}{v3,v4}
\fmf{photon, tension=.5, left}{v1,v2}
\fmf{photon, tension=.5, left}{v3,v4}
\fmfv{d.shape=circle, d.size=3thick}{v1,v2,v3,v4}
\end{fmfgraph*}
} \, + 
\parbox[b][1cm][b]{30mm}{
\begin{fmfgraph*}(70,50)
\fmfbottom{v1,v4}
\fmf{plain}{v1,v2}
\fmf{dashes, tension=2}{v2,v3}
\fmf{plain}{v3,v4}
\fmf{photon, tension=.5, left}{v1,v2}
\fmf{photon, tension=.5, left}{v3,v4}
\fmfv{d.shape=circle, d.size=3thick}{v1,v4}
\end{fmfgraph*}
}
\end{gather}
\end{fmffile}
\caption{Diagrammatic expression for Eq.~(\ref{eq:GSGSG})}
\label{fig:diag_5}
\end{figure} 

If we want the imaginary part of the self energy proportional to
$z^{(D-2)/2}$ we have to consider the term \[\Lambda(\vn p,
z)\equiv\left[\risz\right]^{-1}\sum_{i=1}^{18}L_i^{(2)}\left[\risz\right]^{-1}\]
and the diagrams in Fig. \ref{fig:diag_5}. When we calculate this
contribution we have to set to zero the momentum carried by one
internal propagator $G_0$ so that the contribution coming from
$\Lambda(\vn p, z)$ does not matter. We have to calculate only the
term coming from the Dyson resummation so that the imaginary part of
the self energy at two loops is given by
\begin{fmffile}{diag7}
\begin{gather}
\tilde{\textrm{Im}}\lim_{\epsilon\to 0^+}\Sigma^{(2)}(\vn p, z+i\epsilon)=-2\left[\tilde{\textrm{Im}}\lim_{\epsilon\to 0^+}\Sigma^{(1)}(\vn p, z+i\epsilon)\right]\cdot\\ \cdot \ \ \parbox[b][1cm][b]{20mm}{
\begin{fmfgraph*}(60,30)
\fmfbottom{i1,o1}
\fmf{plain}{i1,v1}
\fmf{dashes, tension=2}{v1,o1}
\fmf{photon, tension=.5, left}{i1,v1}
\end{fmfgraph*}
}-2\Sigma^{(1)}(\vn p, z)\tilde{\textrm{Im}}\lim_{\epsilon\to 0^+}\Biggl[\ \parbox[b][1cm][b]{20mm}{
\begin{fmfgraph*}(60,30)
\fmfbottom{i1,o1}
\fmf{plain}{i1,v1}
\fmf{dashes, tension=2}{v1,o1}
\fmf{photon, tension=.5, left}{i1,v1}
\end{fmfgraph*}
}\ \Biggr]+\\+\ \tilde{\textrm{Im}}\lim_{\epsilon\to 0^+}\Biggl[\,\parbox[b][1cm][b]{28mm}{
\begin{fmfgraph*}(80,30)
\fmfbottom{i1,o1}
\fmf{dashes, tension=2}{i1,v1}
\fmf{plain}{v1,v2}
\fmf{dashes, tension=2}{v2,v3}
\fmf{plain}{v3,v4}
\fmf{dashes, tensio=2}{v4,o1}
\fmf{photon, tension=.5, left}{v1,v2}
\fmf{photon, tension=.5, left}{v3,v4}
\end{fmfgraph*}
}\Biggr]=\\=-2\Sigma^{(1)}(\vn p, z)\tilde{\textrm{Im}}\lim_{\epsilon\to 0^+}\Biggl[\ \parbox[b][1cm][b]{20mm}{
\begin{fmfgraph*}(60,30)
\fmfbottom{i1,o1}
\fmf{plain}{i1,v1}
\fmf{dashes, tension=2}{v1,o1}
\fmf{photon, tension=.5, left}{i1,v1}
\end{fmfgraph*}
}\ \Biggr]\propto -2 z^{(D-2)/2}V(\vn p,\vn p)\Sigma^{(1)}(\vn p,z)\:.
\end{gather}
\end{fmffile} At this point we can give also the analytical argument
\begin{gather}
\tilde{\text{Im}} \lim_{\epsilon\to 0} \Sigma^{(2)}(\vn p,z+i\epsilon)=\\=-\tilde{\text{Im}}\lim_{\epsilon\to 0}\left[\Sigma^{(1)}(\vn p,z+i\epsilon)G_0(\vn p,z+i\epsilon)\Sigma^{(1)}(\vn p,z+i\epsilon)\right]\propto\\ \propto -2V(\vn p,\vn p)\Sigma^{(1)}(\vn p,z)
\end{gather} where we have used the fact that
\begin{equation}
\Sigma(\vn 0,\,z)=0\:.
\end{equation} On the same line we can give the imaginary part proportional to $z^{(D-2)/2}$ at three loops because this contribution comes from the Dyson resummation of one and two loops self energy:
\begin{gather}
\tilde{\text{Im}} \lim_{\epsilon\to 0} \Sigma^{(3)}(\vn p,z+i\epsilon)= -\tilde{\text{Im}} \lim_{\epsilon\to 0}\left[\Sigma^{(1)}G_0\Sigma^{(1)}G_0\Sigma^{(1)}+2\Sigma^{(1)}G_0\Sigma^{(2)}\right]\propto\\
\propto z^{(D-2)/2}\left(3\left[\Sigma^{(1)}\right]^2-2V(\vn p,\vn p)\Sigma^{(2)}\right)\:.
\end{gather} At this point we can give a general expression for the imaginary part proportional to $z^{(D-2)/2}$ at any perturbative order:
\begin{gather}
\tilde{\text{Im}} \lim_{\epsilon\to 0} \Sigma^{(n)}(\vn p,z+i\epsilon)=\\=-\tilde{\text{Im}} \lim_{\epsilon\to 0}\left[
\left( \sum_{k=2}^n \sum_{\underset{\sum i_\sigma=n;\ i_\sigma<n}{i_1,\ldots,i_k}}\prod_{\alpha=1}^k \left(\Sigma^{(i_\alpha)}(\vn p, z+i\epsilon)G_0(\vn p, z+i\epsilon)\right)\right)\right.\cdot \\ \cdot \left.[G_0(\vn p, z+i\epsilon)]^{-1}
\right]\:.\label{eq:non_pert}
\end{gather} From this expression we can prove by induction that the imaginary part of the self energy proportional to $z^{(D-2)/2}$ cannot appear at any order in perturbation theory. In fact we have seen that it does not appear at one and two loop so we can prove that if it does not appear up to $n$ loops it does not appear to $n+1$ loops too.  We can see that
\begin{gather}
\tilde{\text{Im}} \lim_{\epsilon\to 0} \Sigma^{(k)}(\vn p,z+i\epsilon)G_0(\vn p, z+i\epsilon)\propto z^{(D-2)/2}p^\gamma
\end{gather}where $\gamma\geq2$ and where we have showed only the term at lowest order in $z$. Moreover we have
\begin{gather}
\Sigma^{(k)}(\vn p,z)\risz\sim 1 + \mathcal O(p^2)
\end{gather}where we have neglected the higher order in $z$. It follows that the generic term in (\ref{eq:non_pert}) is of order
\begin{equation}
z^{(D-2)/2}p^{\beta}
\end{equation}with $\beta\geq 4$ because $[\risz]^{-1}\sim p^2$. This completes the proof.

\section{Conclusions}

In conclusion, we have given a detailed description of the
perturbative high-density computation of the resolvent (and in
particular the density of states) of \acl{ERM} within two different
formalisms. The combinatorial formalism of
sec.~\ref{SECT:COMBINATORIAL} results in fewer diagrams and is
probably more convenient when the goal is to obtain an expression of
the self-energy at a given order. On the other hand, the
field-theoretic formalism (sec.~\ref{FIELD}), though producing a
higher number of diagrams, has allowed us to analyze the $p\to0$
behavior at all orders in perturbation theory. This analysis shows
that the immaginary part of the self-energy in the limit of small
momenta (which controls the width of the Brillouin peak of the dynamic
structure factor) has, in contrast to previous claims
\cite{ERM1,ERM2,ERM3,ERM4,SCHIRMACHER}, the structure
\begin{equation}
-\text{Im}\,\Sigma(\lambda,p) = {\cal B}\lambda^{\frac{D-2}{2}} p^4 + {\cal C}
\lambda^\frac{D}{2}\frac{ p^2}{c^2} + \ldots\,,
\end{equation}
where ${\cal C},{\cal B}>0$ are amplitudes, and $c$ is the speed of
sound.  This implies in particular a $p^4$ scaling for the Brillouin
peak width, but it also shows that the structure of the theory is more
complex than in the case of scattering from lattice
models~\cite{DIS-CRYSTAL}.

\section{aknowledgements}
We were partly supported by MICINN (Spain) through Research Contract
Nos. FIS2009-12648-C03-01 (VMM and PV) and FIS2008-01323 (PV).

\appendix

\section{Proof of Eq.~(\ref{eq:delta-rho-k-points})}

\label{AP:PROOF}

The proof proceeds by induction. First note that the explicit
computation in Eqs.~(\ref{eq:delta-rho-1-points}) and
~(\ref{eq:delta-rho-2-points}) already implies that
Eq.~\ref{eq:delta-rho-k-points} holds for $k=1$ and $k=2$.

The cornerstone of the proof is a general result for the $k$-point
correlation functions of $\rho$ (rather than $\delta\rho$). The
sought correlation function, in the thermodynamic limit, is
\begin{equation}\label{eq:rho-k-points}
\overline{\rho(\vn{y}_1)
  \rho(\vn{y}_2)\ldots\rho(\vn{y}_k)}=1+\sum_{\omega\in {\cal
    P}^{(k)},\left\Vert\omega\right\Vert<k}
\frac{1}{\rho^{k-\left\Vert\omega\right\Vert}}\left[\prod_{l=1,\left\Vert\Omega_{l,\omega}\right\Vert>1}^{\left\Vert\omega\right\Vert}\left(\prod_{r=1}^{\left\Vert\Omega_{l,\omega}\right\Vert-1}
  \delta(\vn{y}_{\alpha_r^{(l,\omega)}}-\vn{y}_{\alpha_{r+1}^{(l,\omega)}})\right)\right].
\end{equation}
Eq.~(\ref{eq:rho-k-points}) looks very similar to Eq.~(\ref{eq:delta-rho-k-points}),
yet we note the following crucial differences:
\begin{itemize}
\item The partitions $\omega$ belong to ${\cal P}^{(k)}$ rather than
  to the restricted set ${\cal H}^{(k)}$. In particular, the term
  equal to 1 in Eq.~(\ref{eq:rho-k-points}) follows from the only
  partition $\omega$ with $\left\Vert\omega\right\Vert=k$, namely
  $\{\,\{1\},\{3\},\ldots,\{k\}\,\}$, which obviously does not belong
  to ${\cal H}^{(k)}$.
\item In the innermost product in Eq.~(\ref{eq:rho-k-points}), a subset
  $\Omega_{l,\omega}$ with just one element,
  $\left\Vert\Omega_{l,\omega}\right\Vert=1$, merely contributes a factor of
  one. Hence, for all practical purposes, such a subset
  $\Omega_{l,\omega}$ can be ignored.
\end{itemize}

To establish Eq.~(\ref{eq:rho-k-points}), first note that
\begin{equation}
\overline{\rho(\vn{y}_1)
  \rho(\vn{y}_2)\ldots\rho(\vn{y}_k)}=\frac{1}{\rho^k}\sum_{i_1,i_2,\ldots,i_k=1}^N\,
\overline{\delta(\vn{y}_1-\vn{x}_{i_1}) \delta(\vn{y}_1-\vn{x}_{i_2})\ldots
  \delta(\vn{y}_k-\vn{x}_{i_k})},
\end{equation}
where the average is taken with respect to the flat probability measure,
$$\frac{\prod_{i=1}^N \mathrm{d}^D \vn{x}_i}{V^N}.$$ Now, for a
given assignment of the $k$ particle labels $i_1$, $i_2$,\ldots,$i_k$,
we declare that all terms with a coinciding particle label $i_r$ form
a subset $\Omega_{l,\omega}$. It is then obvious that every assignment of
the $k$ particle labels $i_1$, $i_2$,\ldots,$i_k$ defines a partition
$\omega$ in ${\cal P}^{(k)}$. Furthermore, a little reflection shows
that all possible partitions in ${\cal P}^{(k)}$ can be obtained in
this way.  Eq.~(\ref{eq:rho-k-points}) follows from the following three
facts about a given partition $\omega$:
\begin{enumerate}
\item There are $N(N-1)\ldots(N- N_{\left\Vert\omega\right\Vert}$) possible
  assignments of the $k$ particle labels $i_1$, $i_2$,\ldots,$i_k$ that
  yield the partition $\omega$ (you are given $N$ choices for the
  particle that appears in the subset $\Omega_{1,\omega}$, $N-1$ for
  that appearing in $\Omega_{2,\omega}$, and so forth).

\item A subset with a single element, $\left\Vert\Omega_{l,\omega}\right\Vert=1$,
contributes a factor $1/V$.

\item A subset with more than one element, $\left\Vert\Omega_{l,\omega}\right\Vert>1$,
contributes a factor 
$$\frac{1}{V}
\prod_{r=1}^{\left\Vert\Omega_{l,\omega}\right\Vert-1} \delta(\vn{y}_{\alpha_r^{(l,\omega)}}-\vn{y}_{\alpha_{r+1}^{(l,\omega)}})\,.$$
\end{enumerate}

Now consider a partition $\omega$ that belongs to ${\cal P}^{(k)}$
but does {\em not} belong to ${\cal H}^{(k)}$. Imagine that
$\omega$ contains $k-s$ subsets $\Omega_{l,\omega}$ with just one
element. The values that $s$ can take are $s=0,2,3,4,\ldots, k-1$. 
We are not interested in the trivial case $s=0$, that corresponds to
the partition $\{\,\{1\},\{3\},\ldots,\{k\}\,\}$. Hence, for $s>0$,
we simply erase from $\omega$ all the $k-s$ subsets
$\Omega_{l,\omega}$ with $\left\Vert\Omega_{l,s}\right\Vert=1$. The $s$ integers
$$\Lambda=\{\beta_1^\Lambda,\beta_2^\Lambda,\ldots,\beta_s^\Lambda\}\,,$$
that belong to the remaining $\Omega_{l,\omega}$, form the {\em irreducible}
set $\Lambda$ associated to the partition $\omega$. The list of the
$\Omega_{l,\omega}$ with $\left\Vert\Omega_{l,\omega}\right\Vert>1$ provides
a partition $\tilde\omega$ of $\Lambda$, that obviously belongs to
${\cal H}^{(s,\Lambda)}$. Furthermore, $\left\Vert\omega\right\Vert=
k-s+\left\Vert\tilde\omega\right\Vert$, so we have for the prefactor in 
Eq.~(\ref{eq:rho-k-points}) that
$$\frac{1}{\rho^{k-\left\Vert\omega\right\Vert}}=\frac{1}{\rho^{s-\left\Vert\tilde\omega\right\Vert}}.$$
Hence, since $s<k$, the induction hypothesis implies that the added
contribution in Eq.~(\ref{eq:rho-k-points}) of all the partitions
sharing the same irreducible set, $\Lambda$, is
$$\overline{\delta\rho(\vn{y}_{\beta_1^\Lambda})  \delta\rho(\vn{y}_{\beta_2^\Lambda})\ldots\delta\rho(\vn{y}_{\beta_s^\Lambda})}\,.$$
At this point, we may rewrite Eq.~(\ref{eq:rho-k-points}) as
\begin{equation}
\begin{split}
\overline{\rho(\vn{y}_1)
\rho(\vn{y}_2)\ldots\rho(\vn{y}_k)}&=1+\sum_{\omega\in {\cal H}^{(k)}}
\frac{1}{\rho^{k-\left\Vert\omega\right\Vert}}\left[\prod_{l=1}^{\left\Vert\omega\right\Vert}\left(\prod_{r=1}^{\left\Vert\Omega_{l,\omega}\right\Vert-1} \delta(\vn{y}_{\alpha_r^{(l,\omega)}}-\vn{y}_{\alpha_{r+1}^{(l,\omega)}})\right)\right]+\\
& \quad \sum_{s=2}^{k-1}\ \sum_{\Lambda=\{\beta_1^{\Lambda},\beta_1^{\Lambda},\ldots,\beta_s^{\Lambda}\}}\,\overline{\delta\rho(\vn{y}_{\beta_1^\Lambda})  \delta\rho(\vn{y}_{\beta_2^\Lambda})\ldots\delta\rho(\vn{y}_{\beta_s^\Lambda})}\,.\label{eq:app-forma-1}
\end{split}
\end{equation}
We finally note that, if one writes $\rho(\vn{y}_r)=1+
\delta\rho(\vn{y}_r)$,
\begin{equation}
\begin{split}
\overline{\rho(\vn{y}_1)
\rho(\vn{y}_2)\ldots\rho(\vn{y}_k)}&=1+
\overline{\delta\rho(\vn{y}_1)
\delta\rho(\vn{y}_2)\ldots\delta\rho(\vn{y}_k)}
+\\
&\quad \sum_{s=2}^{k-1}\
\sum_{\Lambda=\{\beta_1^{\Lambda},\beta_1^{\Lambda},\ldots,\beta_s^{\Lambda}\}}\,\overline{\delta\rho(\vn{y}_{\beta_1^\Lambda})
  \delta\rho(\vn{y}_{\beta_2^\Lambda})\ldots\delta\rho(\vn{y}_{\beta_s^\Lambda})}.
\label{eq:app-forma-2}
\end{split}
\end{equation}
Comparison of Eqs.~(\ref{eq:app-forma-1}) and~(\ref{eq:app-forma-2})
completes the proof.

\end{document}